\documentclass[fleqn,usenatbib]{mnras}

\usepackage[T1]{fontenc}

\DeclareRobustCommand{\VAN}[3]{#2}
\let\VANthebibliography\thebibliography
\def\thebibliography{\DeclareRobustCommand{\VAN}[3]{##3}\VANthebibliography}

\usepackage{graphicx}	
\usepackage{amsmath}	

\usepackage{newtxtext,newtxmath}

\newcommand\Msun{\,\rmn{M}_{\sun}}

\newcommand\Gyr{\,\rmn{Gyr}}

\newcommand\kpc{\,\rmn{kpc}}

\newcommand\FeH{\lbrack \rmn{Fe}/\rmn{H} \rbrack}

\newcommand{\ts}{\textsuperscript}

\newcommand{\Mcstar}{M_\rmn{c,\ast}}

\title[E-MOSAICS GC metallicity distributions]{Globular cluster metallicity distributions in the E-MOSAICS simulations}

\author[J. Pfeffer et al.]{
Joel Pfeffer$^{1}$\thanks{E-mail: joel.pfeffer@icrar.org (JP)},
J. M. Diederik Kruijssen$^{2}$,
Nate Bastian$^{3,4}$,
Robert A. Crain$^{5}$,
Sebastian Trujillo-Gomez$^{6}$
\\
$^{1}$International Centre for Radio Astronomy Research (ICRAR), M468, University of Western Australia, 35 Stirling Hwy, Crawley, WA 6009, Australia\\
$^{2}$Cosmic Origins Of Life (COOL) Research DAO, coolresearch.io\\
$^{3}$Donostia International Physics Center (DIPC), Paseo Manuel de Lardizabal, 4, E-20018 Donostia-San Sebasti\'{a}n, Guipuzkoa, Spain\\
$^{4}$IKERBASQUE, Basque Foundation for Science, E-48013 Bilbao, Spain\\
$^{5}$Astrophysics Research Institute, Liverpool John Moores University, 146 Brownlow Hill, Liverpool L3 5RF, UK\\
$^{6}$Astronomisches Rechen-Institut, Zentrum f\"{u}r Astronomie der Universit\"{a}t Heidelberg, M\"{o}nchhofstra\ss e 12-14, 69120 Heidelberg, Germany
}

\date{Accepted 2022 December 23. Received 2022 December 14; in original form 2022 October 01}

\pubyear{2022}

\begin{document}
\label{firstpage}
\pagerange{\pageref{firstpage}--\pageref{lastpage}}
\maketitle

\begin{abstract}
The metallicity distributions of globular cluster (GC) systems in galaxies are a critical test of any GC formation scenario.
In this work, we investigate the predicted GC metallicity distributions of galaxies in the MOdelling Star cluster population Assembly In Cosmological Simulations within EAGLE (E-MOSAICS) simulation of a representative cosmological volume ($L = 34.4$~comoving~Mpc).
We find that the predicted GC metallicity distributions and median metallicities from the fiducial E-MOSAICS GC formation model agree well the observed distributions, except for galaxies with masses $M_\ast \sim 2 \times 10^{10} \Msun$, which contain an overabundance of metal-rich GCs.
The predicted fraction of galaxies with bimodal GC metallicity distributions ($37 \pm 2$ per cent in total; $45 \pm 7$ per cent for $M_\ast > 10^{10.5} \Msun$) is in good agreement with observed fractions ($44^{+10}_{-9}$ per cent), as are the mean metallicities of the metal-poor and metal-rich peaks.
We show that, for massive galaxies ($M_\ast > 10^{10} \Msun$), bimodal GC distributions primarily occur as a result of cluster disruption from initially-unimodal distributions, rather than as a result of cluster formation processes.
Based on the distribution of field stars with GC-like abundances in the Milky Way, we suggest that the bimodal GC metallicity distribution of Milky Way GCs also occurred as a result of cluster disruption, rather than formation processes.
We conclude that separate formation processes are not required to explain metal-poor and metal-rich GCs, and that GCs can be considered as the surviving analogues of young massive star clusters that are readily observed to form in the local Universe today.
\end{abstract}

\begin{keywords}
globular clusters: general -- galaxies: star clusters: general -- galaxies: formation -- methods: numerical
\end{keywords}



\section{Introduction}
\label{sec:intro}

Globular clusters (GCs) are one of the most common types of stellar system in the Universe, with most galaxies with stellar masses larger than $10^9 \Msun$ hosting GC populations \citep{Harris_91, Brodie_and_Strader_06}.
However, their formation mechanism is still under debate \citep[see][for recent reviews]{Kruijssen_14, Forbes_et_al_18}, with proposed scenarios broadly falling into two classes: special conditions for GC formation existed in the Early Universe \citep[e.g.][]{Peebles_and_Dicke_68, Peebles_84, Fall_and_Rees_85, Rosenblatt_Faber_and_Blumenthal_88, Trenti_Padoan_and_Jimenez_15, Mandelker_et_al_18, Madau_et_al_20}; or, a universal formation mechanism explains both GCs and young star clusters \citep[e.g.][]{Ashman_and_Zepf_92, Harris_and_Pudritz_94, Kravtsov_and_Gnedin_05, Kruijssen_15, Li_et_al_17, P18, Ma_et_al_20, Reina-Campos_et_al_22c, Grudic_et_al_23}.

One of the most peculiar properties of the GC systems of massive galaxies is their apparent colour and/or metallicity bimodality.
Since the discovery of metallicity bimodality in the Milky Way's GC population \citep{Harris_and_Canterna_79, Freeman_and_Norris_81, Zinn_85}, it has been a major focus in studies of extragalactic GC systems \citep[for reviews, see][]{Ashman_and_Zepf_98, Harris_01, Brodie_and_Strader_06}.
Colour bimodality (or multi-modality) appears commonly in massive galaxies \citep[e.g.][]{Gebhardt_and_Kissler-Patig_99, Larsen_et_al_01}, with a contribution of red GCs that decreases towards fainter galaxies \citep{Peng_et_al_06}, though the interpretation as two distinct populations becomes complicated where internal colour dispersions (inferred through Gaussian fits) become large \citep{Harris_et_al_17}\footnote{A non-bimodal and non-Gaussian distribution can of course be better fit by two (or more) Gaussians than one, see \citet{Larsen_et_al_01} and \citet{Muratov_and_Gnedin_10} for further discussions.}. 
The two populations also often \citep[e.g.][]{Zinn_85, Arnold_et_al_11}, though not always \citep[e.g.][]{Richtler_et_al_04, Dolfi_et_al_21}, show evidence of distinct kinematic signatures, providing further evidence for being distinct populations.

Due to non-linear colour-metallicity relations of GC systems, colour bimodality does not necessarily imply metallicity bimodality \citep{Peng_et_al_06, Yoon_Yi_and_Lee_06, Conroy_Gunn_and_White_09, Usher_et_al_12}.
The GC colour-metallicity relations may also vary from galaxy to galaxy \citep{Usher_et_al_15, Sesto_et_al_18, Villaume_et_al_19}, possibly caused by age or abundance differences, making the inference of metallicity distributions from colour distributions less than straightforward.
Therefore, spectroscopic studies are necessary to confirm the intrinsic metallicity distributions of GC populations.

The more expensive nature of spectroscopy, compared to photometry, means that fewer systematic spectroscopic studies have been undertaken on the metallicity distributions of the GC populations of galaxies. 
This is especially the case for typical star-forming galaxies, since most studies focus on quiescent early-type galaxies, where extinction and reddening from dust are less of an issue.
While NGC 3115 shows one of the best examples of a bimodal GC metallicity distribution \citep{Brodie_et_al_12}, other galaxies often show more diverse distributions \citep{Beasley_et_al_08, Usher_et_al_12, Fahrion_et_al_20b}.
In the Local Group, the GC metallicity distribution of M31 appears to be broad and unimodal, in contrast with the bimodal distribution in the Milky Way \citep{Caldwell_et_al_11}.
Thus, though the bimodality of GC colour distributions may be common, this is not necessarily a universal feature of GC metallicity distributions. 
Any complete scenario for GC formation should therefore explain the origin of the diversity of GC metallicity distributions.

Many scenarios for GC formation have been developed and interpreted through the lens of bimodality. 
Scenarios often invoke separate formation mechanisms and/or truncated formation epochs in order to explain metal-poor and metal-rich GC populations \citep[e.g.][]{Zepf_and_Ashman_93, Forbes_Brodie_and_Grillmair_97, Beasley_et_al_02, Griffen_et_al_10}. 
However a single, continuous formation mechanism may also produce GC bimodality \citep{Kravtsov_and_Gnedin_05, Bekki_et_al_08, Muratov_and_Gnedin_10, Li_and_Gnedin_14, Li_and_Gnedin_19, Kruijssen_15, Choksi_Gnedin_and_Li_18, Keller_et_al_20}. 
Regardless of the actual formation mechanism of GCs, it is thought to be a natural outcome of hierarchical galaxy formation (at least in massive galaxies) that the metal-rich and metal-poor GCs largely represent in-situ and ex-situ formation within their host galaxy, respectively \citep{Cote_Marzke_and_West_98, Hilker_Infante_and_Richtler_99, Tonini_13, Katz_and_Ricotti_14, Renaud_Agertz_and_Gieles_17, Forbes_and_Remus_18, El-Badry_et_al_19, K19a, K19b}.
In this way, the metallicity distributions of GC systems may be intimately tied to the formation and assembly history of their host galaxy, with differing assembly histories resulting in differing GC metallicity distributions \citep[e.g.][]{Tonini_13, K19a}.
Often neglected in studies of GC metallicity distributions, cluster mass loss and disruption may also significantly alter the distributions. For instance, \citet{Kruijssen_15} showed that mass-loss mechanisms are expected to be more effective for higher metallicity GCs. 
Therefore, present-day distributions could be significantly different from the initial distributions, complicating the inference of formation scenarios from present-day GC population properties.

Closely related to GC metallicity distributions are the specific frequencies of GCs (the number of GCs relative to the total luminosity, mass or number of field stars) as a function of metallicity in galaxies. 
Observations show that the specific frequency is a strong function of metallicity and decreases with increasing metallicity; i.e., more stars are found in GCs at low metallicities ($\FeH \sim -2$) than at higher metallicities \citep{Durrell_Harris_and_Pritchet_01, Harris_and_Harris_02, Beasley_et_al_08, Gratton_Carretta_and_Bragaglia_12, Kruijssen_15, Lamers_et_al_17}.
This relation could be imparted either at formation (more efficient formation of low metallicity clusters), by cluster disruption (higher mass-loss rates for higher metallicity clusters), or some combination of both, and presents an important diagnostic for models of GC system formation and evolution \citep[see][for a comprehensive discussion]{Lamers_et_al_17}.

In this work, we investigate the metallicity distributions and metallicity-dependent specific frequencies of GCs in the MOdelling Star cluster population Assembly In Cosmological Simulations
within EAGLE \citep[E-MOSAICS,][]{P18, K19a} simulation of galaxy and star cluster formation within a periodic cosmological volume of side length $L = 34.4$~comoving~Mpc.
E-MOSAICS couples models for star cluster formation and evolution to the Evolution and Assembly of GaLaxies and their Environments (EAGLE) galaxy formation model \citep{S15, C15}, under the assumption that both young and old star clusters form and evolve following the same physical mechanisms. 
We aim to test whether a common formation mechanism for young and old clusters can explain the diversity of metallicity distributions found for GC systems of observed galaxies.

This paper is organised as follows.
In Section~\ref{sec:methods} we briefly describe the E-MOSAICS model and our GC selection methods for the simulation.
Section~\ref{sec:results} presents the results from this study, including predictions for GC metallicity distributions, specific frequency-metallicity relations and dependence on the GC formation model.
In Section~\ref{sec:discussion} we discuss the origin of bimodal GC metallicity distributions and connection to GC formation scenarios.
Finally, in Section~\ref{sec:summary} we summarise the conclusions of this work.

\section{Methods}
\label{sec:methods}

In this section we briefly describe the E-MOSAICS and EAGLE models, the periodic volume simulation analysed in this work and our selection of GCs from the simulation.

\subsection{E-MOSAICS simulations}
\label{sec:emosaics}

E-MOSAICS \citep{P18, K19a} is a suite of cosmological, hydrodynamical simulations that couple the MOSAICS \citep{Kruijssen_et_al_11, P18} model for star cluster formation and evolution to the EAGLE galaxy formation model \citep{S15, C15}.
The simulations were performed with a highly-modified version of the $N$-body, smooth particle hydrodynamics code \textsc{Gadget 3} \citep{Springel_05} and the EAGLE model includes subgrid routines for radiative cooling \citep{Wiersma_Schaye_and_Smith_09}, star formation \citep{Schaye_and_Dalla_Vecchia_08}, stellar evolution \citep{Wiersma_et_al_09}, the seeding and growth of supermassive black holes \citep{Rosas-Guevara_et_al_15} and feedback from star formation and black hole growth \citep{Booth_and_Schaye_09}.
In EAGLE, the feedback efficiencies for stellar and black hole feedback are calibrated such that simulations of representative galaxy populations reproduce the galaxy stellar mass function, galaxy sizes and black hole masses at $z \approx 0$ \citep{C15}.
Simulations using the EAGLE model have been well studied and shown to broadly reproduce many features of the evolving galaxy population, including the evolution of the galaxy stellar mass function \citep{Furlong_et_al_15} and galaxy sizes \citep{Furlong_et_al_17}, galaxy star formation rates and colours \citep{Furlong_et_al_15, Trayford_et_al_17}, cold gas properties \citep{Lagos_et_al_15, Crain_et_al_17}, galaxy morphologies \citep{Bignone_et_al_20} and (particularly relevant for this work) the galaxy mass-metallicity relation \citep[][with the high-resolution `recalibrated' model used in this work]{S15}.

In the MOSAICS model, star clusters are treated as subgrid components of stellar particles, such that they adopt the properties of their host particle (positions, velocities, ages, abundances). 
Star cluster formation is described by two functions: the fraction of stars formed in bound clusters \citep[i.e.\ the cluster formation efficiency, CFE,][]{Bastian_08} and the shape of the initial cluster mass function (a power law or \citealt{Schechter_76} function, i.e.\ power law with a high mass exponential truncation $\Mcstar$).
For both the power law and Schechter initial mass functions the initial power law index is set to be $-2$, based on observations of young clusters \citep[e.g.][]{Zhang_and_Fall_99, Bik_et_al_03, Gieles_et_al_06b, McCrady_and_Graham_07}.
In each case, the CFE or $\Mcstar$ may be constant or vary with the properties of the local environment at the time of star formation (see below). 
Each newly-formed star particle may (stochastically) form a fraction of its mass in clusters (i.e.\ the CFE times the particle mass) and clusters are sampled stochastically from the initial mass function (such that the subgrid clusters may be more massive than the stellar mass of the host particle).
Thus the total cluster and field star mass is only conserved for an ensemble of star particles \citep[for details of the method, see][]{P18}.
Following formation, star clusters may lose mass by stellar evolution (following the EAGLE model), two-body relaxation depending on the local tidal field strength (\citealt{Lamers_et_al_05b}; \citealt{Kruijssen_et_al_11}; with an additional constant term to account for isolated clusters, following \citealt{Gieles_and_Baumgardt_08}) and tidal shocks from rapidly changing tidal fields \citep{Gnedin_Hernquist_and_Ostriker_99, Prieto_and_Gnedin_08, Kruijssen_et_al_11}.
Dynamical friction is treated in post-processing at every snapshot and assumed to completely remove clusters when the dynamical friction timescale is less than the cluster age \citep[assuming they merge to the centre of their host galaxy, see][]{P18}.

In the E-MOSAICS suite we consider four main variations of the cluster formation model, i.e.\ two variations each (constant or environmentally varying) for the CFE and initial cluster mass function.
Comparison of the four models enables the most critical aspect of cluster formation to be determined for a particular observable.
The environmentally-varying CFE is determined by the \citet{Kruijssen_12} model, which scales as a function of the natal gas pressure such that higher pressures result in higher CFE \citep[see figure 3 in][]{P18}.
The environmentally-varying mass function has an exponential truncation ($\Mcstar$) which varies with local gas and dynamical properties according to the model of \citet{Reina-Campos_and_Kruijssen_17}, such that $\Mcstar$ increases with local gas pressure, except where limited by high Coriolis or centrifugal forces (i.e.\ near the centres of galaxies).
Thus the four cluster formation models are:
\begin{itemize}
\item \textit{Fiducial}: Both the CFE and $\Mcstar$ are environmentally dependent. The fiducial model reproduces observed scaling relations of young star clusters \citep{Pfeffer_et_al_19b}.
\item \textit{CFE only}: Environmentally-varying CFE and pure power-law mass function (i.e.\ $\Mcstar = \infty$).
\item \textit{$\Mcstar$ only}: Constant $\mathrm{CFE} = 0.1$ and environmentally-varying $\Mcstar$.
\item \textit{Constant formation}: Constant $\mathrm{CFE} = 0.1$ and power-law mass function ($\Mcstar = \infty$).
\end{itemize}

The fiducial E-MOSAICS model has previously been shown to produce GC populations consistent with many observed relations, including the fraction of stars contained in GCs \citep{Bastian_et_al_20}, GC system radial distributions \citep{Reina-Campos_et_al_22a}, the high-mass truncation of GC mass functions \citep{Hughes_et_al_22}, the `blue tilt' GC colour distributions \citep{Usher_et_al_18} and the age-metallicity relations of GC systems \citep{K19b, Kruijssen_20, Horta_et_al_21a}.
The alternative formation models (CFE only, $\Mcstar$ only, constant formation) generally fail to simultaneously reproduce particular aspects of star cluster populations \citep{Usher_et_al_18, Pfeffer_et_al_19b, Reina-Campos_et_al_19, Bastian_et_al_20, Hughes_et_al_22}.
As discussed in \citet{P18} and \citet{K19a}, in Milky Way-mass haloes the E-MOSAICS simulations over-predict the number of low-mass and high-metallicity GCs.
This issue is postulated to be a result of a lack of a dense substructure in the interstellar medium (ISM) which may disrupt clusters through tidal shocks, as the EAGLE model does not simulate the cold, dense ISM phase. 
This issue may be remedied in simulations which include models for the cold ISM \citep[e.g.][]{Reina-Campos_et_al_22c}.

In this work, we analyse the E-MOSAICS simulation of a periodic volume with side length $34.4$~comoving~Mpc (first reported in \citealt{Bastian_et_al_20}).
In the simulation, all four cluster formation models were performed in parallel, which is possible as the EAGLE galaxy model is independent of the MOSAICS cluster model.
The simulation adopts the `Recalibrated' EAGLE model parameters \citep{S15} and initially has $1034^3$ dark matter and gas particles with masses of $m_\mathrm{dm} = 1.21 \times 10^6 \Msun$ and $m_\mathrm{b} = 2.26 \times 10^5 \Msun$, respectively.
The simulation adopts parameters consistent with a \citet{Planck_2014_paperXVI} cosmology ($\Omega_\mathrm{m} = 0.307$, $\Omega_\Lambda = 0.693$, $\Omega_\mathrm{b} = 0.04825$, $h = 0.6777$, $\sigma_8 = 0.8288$).
Galaxies (subhaloes) were identified in simulation snapshots by first detecting dark matter structures with the friends-of-friends (FoF) algorithm \citep{Davis_et_al_85} and then identifying bound subhaloes within each FoF group using the \textsc{subfind} \citep{Springel_et_al_01, Dolag_et_al_09} algorithm.
The subhalo containing the particle with the minimum gravitational potential is defined as the central galaxy in each FoF group.

\subsection{Globular cluster selection}
\label{sec:GC_selection}

\begin{figure}
    \includegraphics[width=84mm]{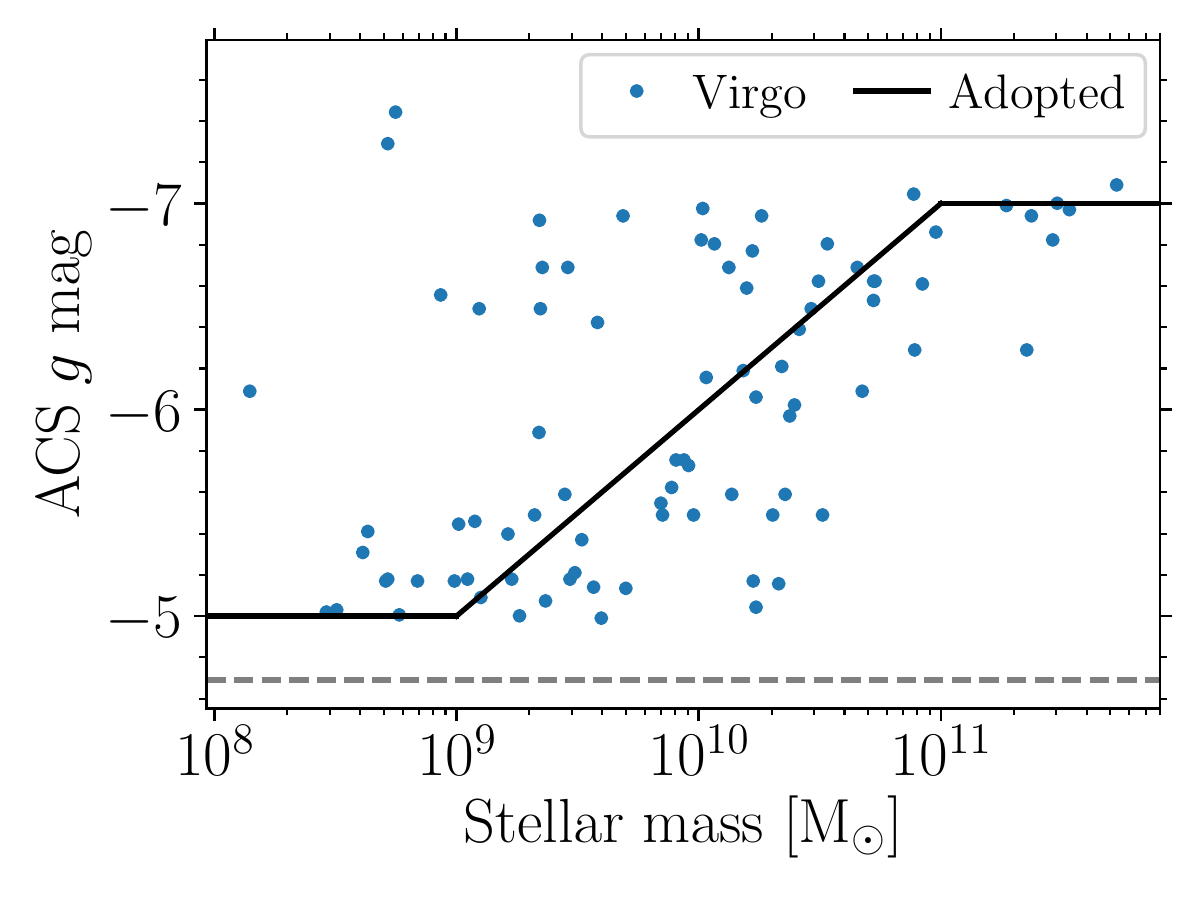}
    \caption{Adopted lower luminosity limits in the ACS $g$ band for GC selection in the simulation as a function of host galaxy mass (black line). The scaling is based on the 90 per cent completeness limits for GCs in galaxies from the ACS Virgo Cluster Survey \citep[shown as blue dots;][]{Jordan_et_al_07_XII}, and approximately accounts for the difficulty in detecting fainter GCs in higher surface brightness galaxies. The grey dashed line shows the lower luminosity limit for GC detection in Virgo galaxies.}
    \label{fig:completeness}
\end{figure}

Given the large variations in GC selection between different observational studies/individual galaxies (e.g.\ due to differing distances of galaxies, background field star surface brightnesses, spectroscopic limitations, etc.), it is difficult to perform systematic comparisons with consistent GC selection.
We base our GC selection on the GC luminosity limits of the ACS Virgo/Fornax surveys.
In particular, we aim to compare predictions from the simulations against observations of the Virgo cluster from \citet{Peng_et_al_06} and the Fornax cluster from \citet{Fahrion_et_al_20a, Fahrion_et_al_20b}.
Additionally, we select GCs with ages $> 2 \Gyr$ (i.e.\ we exclude young star clusters which may undergo significant mass loss through stellar evolution) and metallicities $\FeH > -3$ (due to resolution limits, i.e.\ baryonic particle masses $\approx 2 \times 10^5 \Msun$, very low metallicity particles are poorly sampled in the simulations).

Following table 1 in \citet[][with galaxy stellar masses from \citealt{Peng_et_al_06}]{Jordan_et_al_07_XII}, GCs in $10^9 \Msun$ galaxies in Virgo are $\sim 90$ per cent complete at the lower apparent magnitude limit of $m_g \approx 26.4$ mag.
Virgo galaxies with stellar masses $\approx 10^{11} \Msun$ have 90 per cent GC completeness at $m_g \approx 24.3$ mag.
The difference in distance between the Virgo and Fornax clusters implies mass limits around 40 per cent larger in the Fornax cluster (assuming distance moduli of $31.09$ and $31.51$, respectively, \citealt{Peng_et_al_06, Blakeslee_et_al_09}).
Thus we adopt lower GC luminosity limits in the ACS $g$ band with a floor at $M_g = -5$ and a ceiling at $M_g = -7$, that scales linearly with the logarithmic stellar mass from $M_g = -5$ at $\log(M_\ast / \mathrm{M}_{\sun}) = 9$ to $M_g = -7$ at $\log(M_\ast / \mathrm{M}_{\sun}) = 11$.
We compare our adopted limits with the Virgo galaxy completeness limits in Fig.~\ref{fig:completeness}.

Colours were generated for the simulated GCs using the Flexible Stellar Population Synthesis (\textsc{fsps}) models \citep{Conroy_Gunn_and_White_09, Conroy_and_Gunn_10}, assuming simple stellar populations and using the MILES spectral library \citep{Sanchez-Blazquez_et_al_06}, Padova isochrones \citep{Girardi_et_al_00, Marigo_and_Girardi_07, Marigo_et_al_08}, a \citet{Chabrier_03} initial stellar mass function (as assumed in the EAGLE model) and assuming the default \textsc{fsps} parameters.
Mass-to-light ratios for the GCs were calculated by linearly interpolating the from the grid in ages and total metallicities ($\log Z / \mathrm{Z}_{\sun}$).

\section{Results}
\label{sec:results}

\subsection{Average metallicity distributions}
\label{sec:met_dists}

\begin{figure*}
    \includegraphics[width=\textwidth]{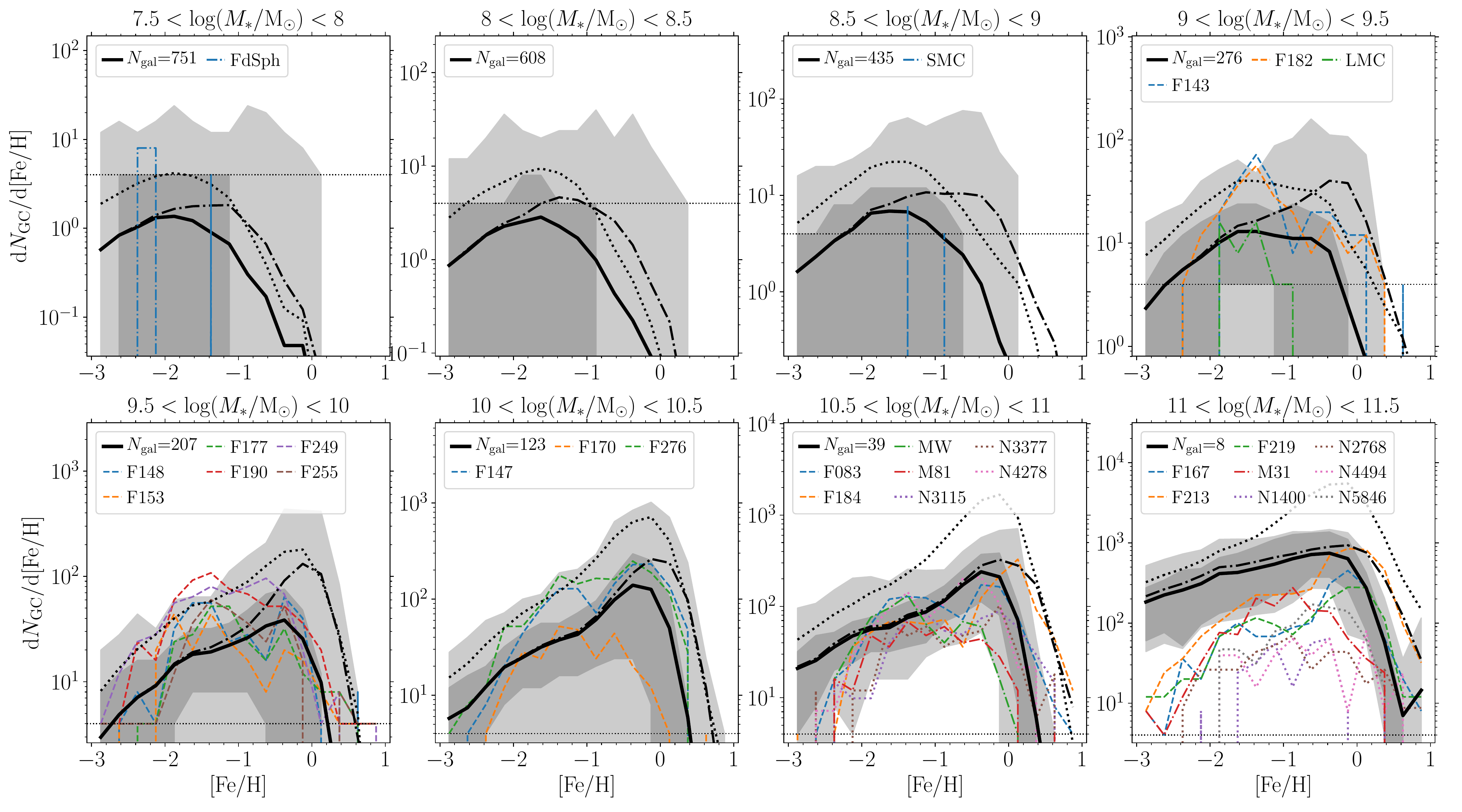}
    \caption{Average GC metallicity distributions in E-MOSAICS galaxies compared with observed galaxies. Each panel shows a different mass range (indicated in titles), from low mass (top left panel) to high mass (bottom right panel). Black lines are GC metallicity distributions averaged over all galaxies in each mass range. Solid and dash-dotted black lines show GC age limits of $>8$ and $>2 \Gyr$, respectively. Black dotted lines show the `initial' GC distributions (for ages $>8 \Gyr$) without dynamical mass loss (initial masses with only stellar evolutionary mass loss applied). The shaded regions show the 16\ts{th}-84\ts{th} (dark grey) and 0\ts{th}-100\ts{th} (light grey) percentiles for all galaxies in each subpanel (GC ages $>8 \Gyr$). For comparison we show galaxies from the Fornax cluster \citep[dashed colour lines;][]{Fahrion_et_al_20b}, the SLUGGS survey \citep[dotted colour lines;][]{Usher_et_al_12} and field galaxies (dash-dotted colour lines; mainly from the Local Group: Milky Way, M31, M81, LMC, SMC, Fornax dSph). Note that for galaxies $>10^{11} \Msun$ the apparent over-abundance of low metallicity GCs in the simulations is due to the ACS Fornax Cluster Survey field of view (see Fig.~\ref{fig:Rcut}). In each panel, the dashed horizontal lines shows the limit of 1 GC per metallicity bin.}
    \label{fig:met_dists}
\end{figure*}

\begin{figure}
    \includegraphics[width=\columnwidth]{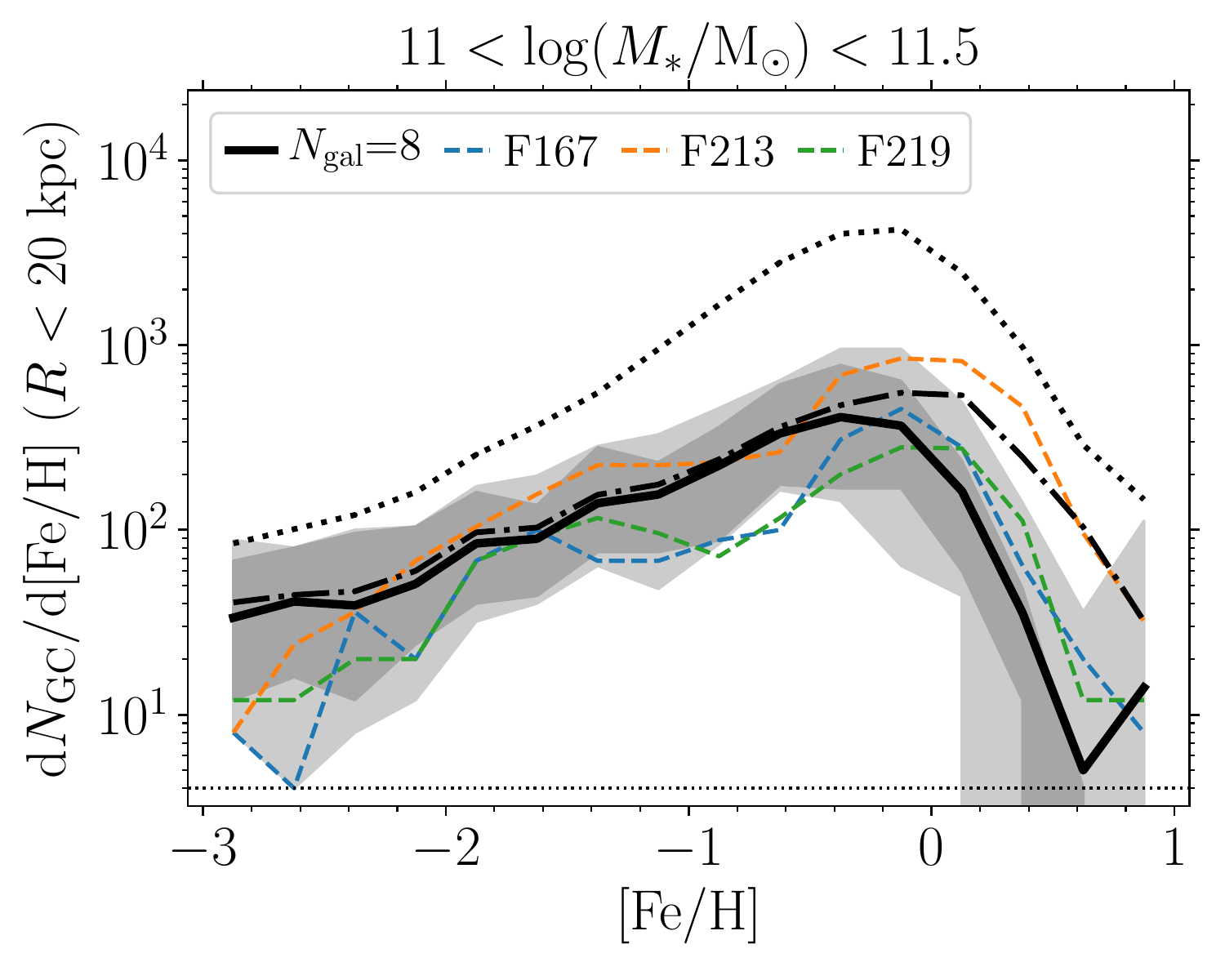}
    \caption{As for Fig.~\ref{fig:met_dists}, but comparing massive ($>10^{11} \Msun$) galaxies with a maximum projected galactocentric radius limit for GCs similar to the ACS Fornax Cluster Survey field of view ($20\kpc$).}
    \label{fig:Rcut}
\end{figure}

In Fig.~\ref{fig:met_dists} we show the average GC metallicity distributions (i.e.\ stacked distributions normalised by number of galaxies) predicted in the E-MOSAICS simulations as a function of galaxy mass.
For the fiducial comparisons we consider GCs with ages $>8 \Gyr$.
This age selection is similar to the youngest Milky Way GCs \citep[e.g.][]{Forbes_and_Bridges_10} and most similar to the old ages of GCs in cluster galaxies \citep[which undergo early star formation quenching, e.g.][]{Gallazzi_et_al_21}.
We also show predictions for ages $>2 \Gyr$ as black dash-dotted lines for reference.
The effect of a younger age limit ($2 \Gyr$) shows the most difference in galaxies with masses $8.5 \lesssim \log(M_\ast / \rmn{M}_{\sun}) < 10$, which is due to galaxy downsizing \citep[i.e.\ lower-mass galaxies form more mass at later times,][]{Bower_et_al_92, Gallazzi_et_al_05}.

In the figure we also compare the simulations against observations from the Fornax cluster \citep{Fahrion_et_al_20a, Fahrion_et_al_20b}, SLUGGS survey \citep{Usher_et_al_12, Usher_et_al_15, Usher_et_al_19} and local field galaxies (Milky Way, \citealt{Harris_96, Harris_10}; M31, \citealt{Caldwell_et_al_11}; M81, \citealt{Nantais_and_Huchra_10}; Large and Small Magellanic Cloud, \citealt{Horta_et_al_21a}; Fornax dwarf spheroidal, \citealt{Larsen_et_al_12}).
Where possible, for Local Group galaxies we also apply the same $>8 \Gyr$ cluster age limit as the simulation (i.e.\ for the Large and Small Magellanic Clouds).
We note that the SLUGGS GC samples are not complete (particularly for the most massive galaxies), and thus only the relative distributions can be compared in these cases.
Additionally, we have not attempted to match the environments or morphologies of the observed galaxies, which vary significantly (i.e.\ from field galaxies to galaxy clusters), and simply compare all galaxies contained in the periodic volume.

In general, the predictions from simulations show a good match to observed GC metallicity distributions.
In most cases, the observed distributions for each galaxy mass are fully contained within the galaxy-to-galaxy scatter of the simulations (grey shaded regions), except for the highest mass galaxies ($M_\ast > 10^{11} \Msun$) and very low metallicity GCs ($\FeH < -2.5$) which we discuss below.
In the $9.5 < \log_{10}(M_\ast / \mathrm{M}_{\sun}) < 10$ mass range GCs with metallicities $\FeH \sim -1.5$ appear under-abundant in simulated galaxies by a factor $\approx 2$.
Potentially, this could be due to inefficient formation of GCs in such galaxies, which we discuss further in Section~\ref{sec:alt-phys}, or simply that our GC selection (Section~\ref{sec:GC_selection}) is too strict in comparison to these Fornax cluster galaxies.
For Milky Way-mass galaxies [$10.5 < \log_{10}(M_\ast / \mathrm{M}_{\sun}) < 11$], the simulations predict a factor $\approx 2.5$ times more high-metallicity GCs ($\FeH \sim -0.5$) than typically observed.
This was previously discussed by \citet{K19a} and suggested to be caused by insufficient GC disruption in galactic discs in the simulations (see also Section~\ref{sec:emosaics}).

In the figure we also compare the `initial' GC metallicity distributions (black dotted lines; ages $>8 \Gyr$), i.e.\ assuming the only cluster mass loss was due to stellar evolution.
On average (individual galaxies may of course differ), for galaxies with $M_\ast < 10^{10} \Msun$ the $z=0$ GC metallicity distributions (solid black lines) are a reasonable reflection of the initial distributions, but typically a factor 3-4 lower due to dynamical mass loss.
However, at higher masses there is a clear dependence of GC `survival' on metallicity (at least above the adopted GC luminosity limit), with surviving high-metallicity ($\FeH \sim 0$) clusters representing a smaller fraction (about one tenth) of the initial population than at lower metallicities.
This effect also appears to increase in higher-mass galaxies.
We will return to this point in Section~\ref{sec:TN-met} when discussing the specific frequency-metallicity relation.

In Fig.~\ref{fig:met_dists} the very massive galaxies ($>10^{11} \Msun$) appear to have too many low metallicity clusters.
In Fig.~\ref{fig:Rcut} we show this is due to the small ACS Fornax Cluster Survey field of view ($202 \times 202$ arcseconds), combined with the radial gradients in GC metallicity.
Adopting a similar radius limit (projected radius of $20 \kpc$), we find excellent agreement with the metallicity distributions of Fornax galaxies.
Potentially, at $\FeH \approx -3$ the number of simulated GCs are still over-predicted by a factor of $\approx 3$ (which is similarly suggested in the $10.5 < \log_{10}(M_\ast / \mathrm{M}_{\sun}) < 11$ panel of Fig.~\ref{fig:met_dists}).
Such particles form in very low mass galaxies \citep[i.e.\ $M_\ast \lesssim 10^6 \Msun$, as inferred in figure 9 of][]{K19a} which are poorly resolved in the simulation (with initial baryonic particle masses of $2.25\times 10^5 \Msun$).
Thus an over-abundance of extremely low-metallicity GCs could stem from the GC formation model performing incorrectly in poorly-resolved galaxies (e.g.\ overestimating $\Mcstar$).

\subsection{Specific frequency-metallicity relationship}
\label{sec:TN-met}

\begin{figure*}
    \includegraphics[width=\textwidth]{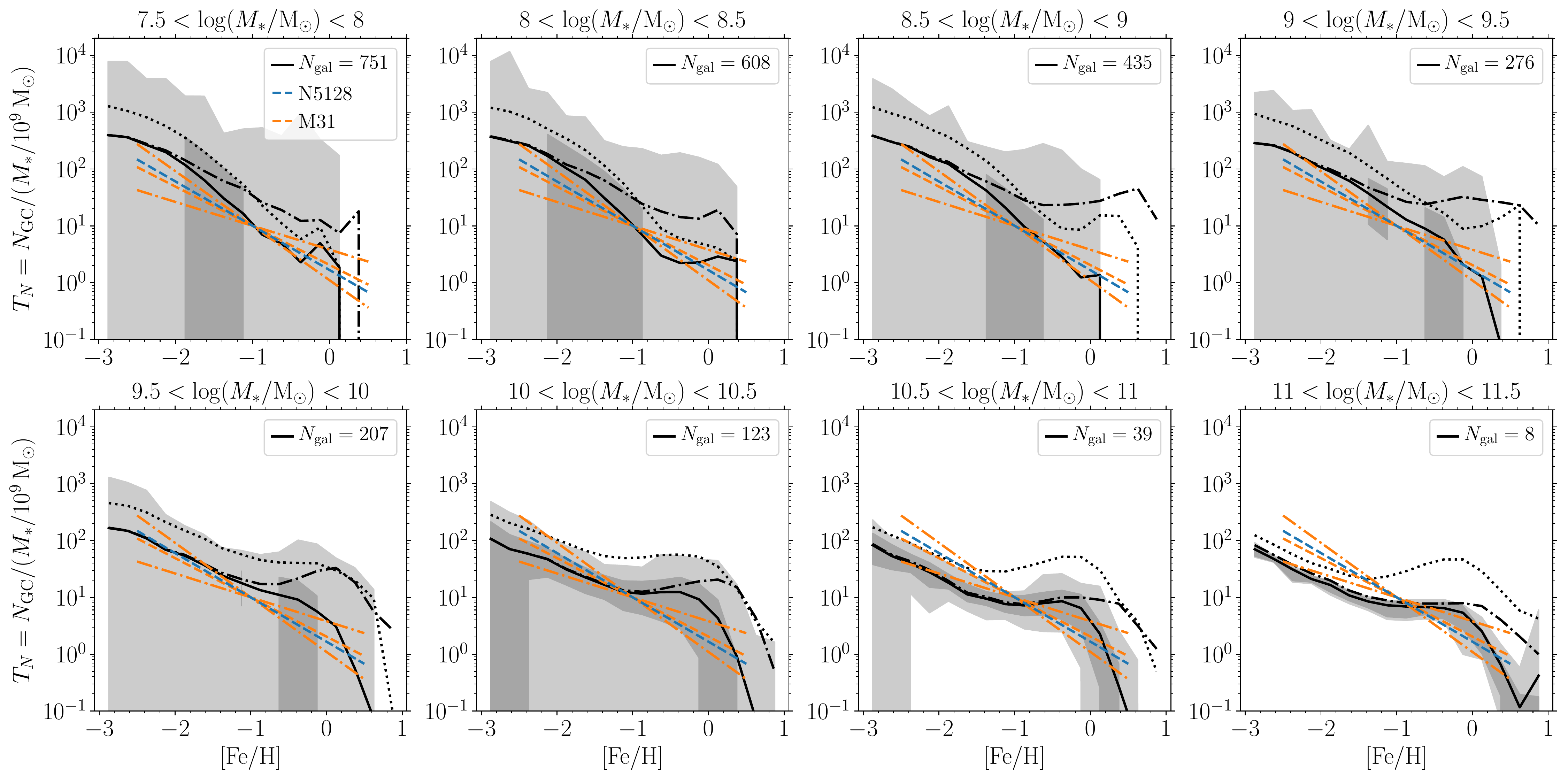}
    \caption{Relationship between GC specific frequency ($T_N$) and metallicity in E-MOSAICS galaxies. Each panel shows a different mass range (indicated in titles), from low mass (top left panel) to high mass (bottom right panel), as in Fig.~\ref{fig:met_dists}. Solid and dash-dotted black lines show age limits $>8 \Gyr$ and $>2 \Gyr$. The dotted black lines show initial distributions for ages $>8 \Gyr$. The shaded regions show the 16\ts{th}-84\ts{th} (dark grey) and 0\ts{th}-100\ts{th} (light grey) percentiles for all galaxies in each subpanel. Coloured dashed lines show the slopes for NGC 5128 and M31 from \citet{Lamers_et_al_17}, normalised to $T_N = 10$ at $\FeH = -1$ \citep[based on figure 1 in][]{Kruijssen_15}, while dash-dotted lines show the slope uncertainties for M31 ($0.69 \pm 0.27$, the variation in slope for NGC 5128 as function of galactocentric distance is also contained within these uncertainties). Absolute $T_N$ depends on the GC luminosity limit ($T_N$ is higher in low-mass galaxies due to the fainter GC luminosity limit), therefore the slope of \citet{Lamers_et_al_17} relations is the relevant comparison.}
    \label{fig:TN}
\end{figure*}

In Fig.~\ref{fig:TN} we compare the relationship between specific frequency (i.e.\ number of GCs per unit galaxy mass; $T_N = N_\mathrm{GC} / (M_\ast / \mathrm{M}_{\sun})$) and metallicity in the E-MOSAICS galaxies.
The simulations predict a strong dependence of $T_N$ on metallicity, with more GCs per unit mass at lower metallicities, as found for observed galaxies \citep{Harris_and_Harris_02, Lamers_et_al_17}.
Across all stellar mass ranges, the slope of the $T_N$-$\FeH$ relations for simulated galaxies (at least for GC ages $>8 \Gyr$) is in reasonable agreement with that found for NGC 5128 and M31 \citep{Lamers_et_al_17}.
The simulations predict slightly flatter slopes for the $T_N$-$\FeH$ relation at higher galaxy masses, though we caution that this result may be strongly affected by the under-disruption of GCs in the simulations.
As the EAGLE model does not resolve the cold, dense phase of the interstellar medium (ISM), mass loss through tidal shocks by dense substructure is under-estimated in the E-MOSAICS simulations, particularly at higher metallicities \citep[see][for further details and discussion]{P18, K19a}.

In galaxies with stellar masses $M_\ast < 10^{10} \Msun$, the $T_N$-$\FeH$ relation (solid line) generally follows the initial relation (dotted line); i.e.\ the $T_N$-$\FeH$ relation is largely set by cluster formation alone.
In the E-MOSAICS model, this increase of $T_N$ towards low metallicities at the time of GC formation is driven by the CFE (fraction of star formation in bound clusters), which tends to increase at lower metallicities \citep{P18}.
This increase is a result of the metallicity-dependent density threshold for star formation implemented in EAGLE, which accounts for the transition from a warm, photoionized interstellar gas to a cold, molecular phase, that is expected to occur at lower densities/pressures at higher metallicities \citep{Schaye_04}.

Towards higher galaxy masses, the initial relations show an increase in $T_N$ at $\FeH \sim 0$.
Cluster mass loss then begins to play a more significant role in setting GC numbers, particularly at $\FeH \gtrsim -1$ (as noted in Section~\ref{sec:met_dists}; see \citealt{Kruijssen_15}), i.e.\ cluster mass loss preferentially affects higher-metallicity GCs in higher-mass galaxies.
Both processes are due to an increase in higher density/pressure gas at higher galaxy masses, resulting in both increased cluster formation (higher CFE) and cluster disruption \citep[i.e.\ the `cruel cradle effect',][]{Kruijssen_et_al_12b}.
We note that in other GC formation models the specific frequency-metallicity relation may have different origins.
For example, in the \citet{Choksi_Gnedin_and_Li_18} model the specific frequency-metallicity relation is set by cluster formation (GC disruption in the model is environmentally independent), while they find a good match to the observed slope of the relation.

\begin{figure*}
    \includegraphics[width=\textwidth]{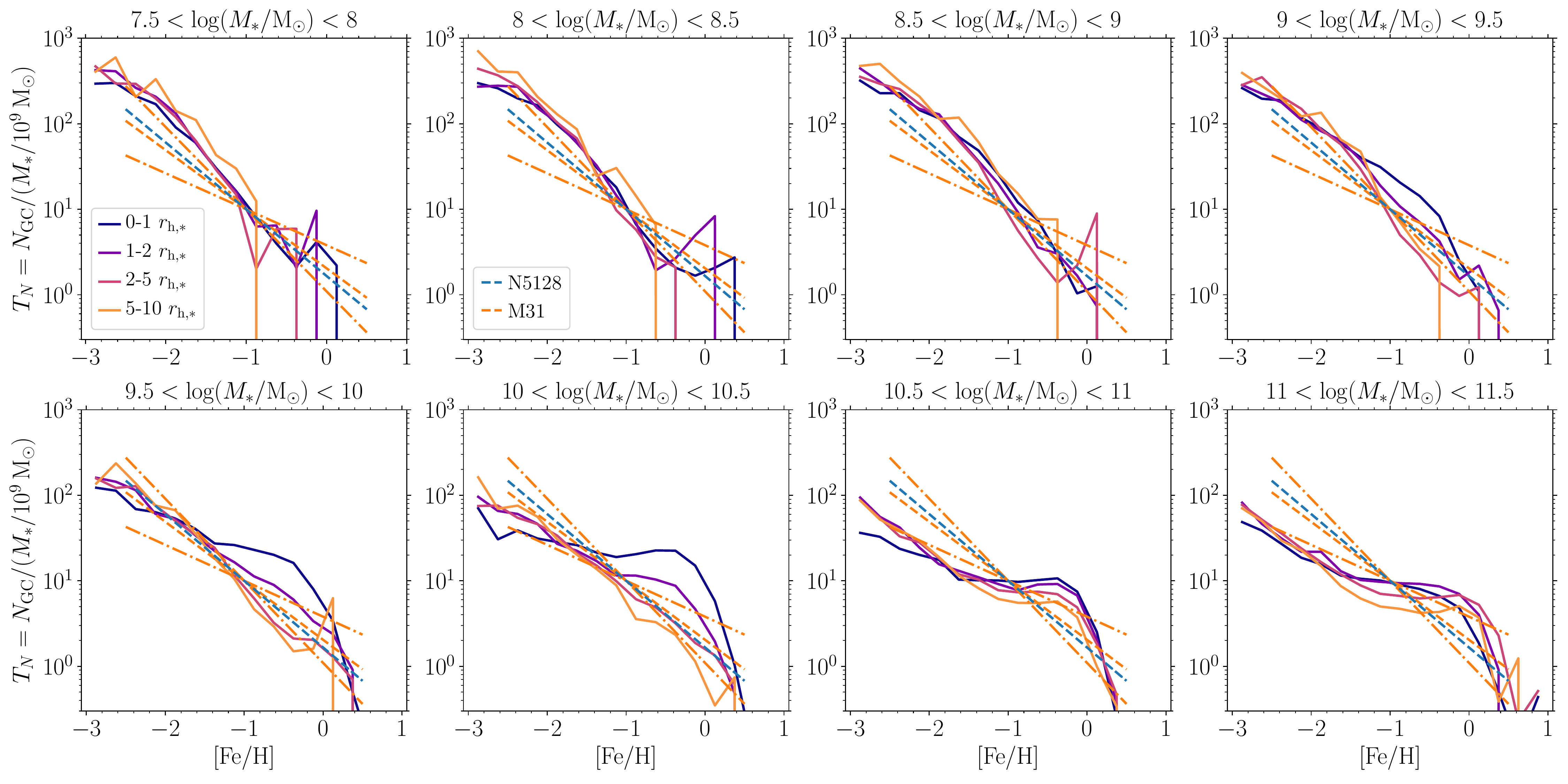}
    \caption{Dependence of the GC specific frequency-metallicity relation on 3D radius in E-MOSAICS galaxies. Within each galaxy, the distances are normalised to the 3D stellar half-mass radius, $r_\mathrm{h,\ast}$, in four radial bins (0-1, 1-2, 2-5 and 5-$10 r_\mathrm{h,\ast}$). Each panel shows the average relations for different galaxy stellar mass ranges (indicated in subpanel titles) as in Figures~\ref{fig:met_dists} and \ref{fig:TN}. For reference, the $\FeH$-$T_N$ relations for M31 and NGC 5128 \citep{Lamers_et_al_17} are shown in each panel, as in Fig.~\ref{fig:TN}.}
    \label{fig:TN_radius}
\end{figure*}

Observations of the $T_N$-$\FeH$ relation are generally restricted to galaxy haloes and exclude the galaxy discs \citep[e.g.\ $R > 6 \kpc$ for NGC 5128 and $R > 12 \kpc$ for M31,][]{Harris_and_Harris_02, Lamers_et_al_17}.
Therefore in Fig.~\ref{fig:TN_radius} we compare the relations as a function of (3D) radius within the simulated galaxies.
To account for different sizes of galaxies, we compare four radial ranges scaled by the 3D stellar half-mass radius (0-1, 1-2, 2-5 and 5-$10 r_\mathrm{h,\ast}$).

For galaxy stellar masses $M_\ast < 10^9 \Msun$ we do not find strong differences in the $T_N$-$\FeH$ relation as a function of radius.
In the range $10^{9}$-$10^{10.5} \Msun$ there are clear radial gradients in $T_N$ at $\FeH \gtrsim -1$, such that $T_N$ is higher at smaller radii.
This is driven by radial gas pressure gradients in galaxies, which results in higher CFE at smaller radii \citep[see figure 4 in][]{P18}, and an ex-situ origin for GCs at larger radii through the accretion of lower mass galaxies \citep[figure 8 in][]{Reina-Campos_et_al_22a}.
At radii $>5 r_\mathrm{h,\ast}$ (i.e.\ the `halo' of the galaxies) the $T_N$-$\FeH$ relations are in good agreement with the observations from \citet{Lamers_et_al_17} due to this (largely) ex-situ origin, despite the under-disruption of GCs at smaller radii.

For galaxy stellar masses $M_\ast > 10^{10.5} \Msun$ we find that $T_N$ at high metallicities ($\FeH \gtrsim -1$) is increased at all radii.
This is due to the increased importance of mergers, particularly major mergers, in the assembly of massive galaxies \citep{Qu_et_al_17}.
As a result, an ex-situ origin for GCs becomes dominant in massive galaxies at nearly all radii \citep{Reina-Campos_et_al_22a,Trujillo-Gomez_et_al_22} and metallicities (Section~\ref{sec:bimod_origin}).
The merging of massive galaxies (with elevated $T_N$ at high $\FeH$ and small radii) then results in results in elevated $T_N$ at high $\FeH$ and large radii in the merger descendant.

Interestingly, comparison of the $T_N$-$\FeH$ relations (Figs.~\ref{fig:TN} and \ref{fig:TN_radius}) and GC metallicity distributions (Figs.~\ref{fig:met_dists} and \ref{fig:Rcut}) for the highest mass galaxies ($M_\ast > 10^{11} \Msun$) suggests that the $T_N$-$\FeH$ relation may differ in Fornax cluster galaxies.
While the simulated $T_N$-$\FeH$ relation is flatter than observed in M31 and NGC 5128 (and similarly, the number of high-metallicity GCs is larger than observed in M31, Fig.~\ref{fig:met_dists}), we find good agreement with the metallicity distributions for Fornax cluster galaxies (Fig.~\ref{fig:Rcut}).
The origin of this difference in the observations is not clear, but plausibly could be due to differences in the formation histories of cluster and field galaxies (such as differing galaxy merger or star-formation quenching histories).
Ideally, future observations should test for an environmental dependence in the $T_N$-$\FeH$ relations of galaxies \citep[which might be possible in the Virgo cluster using observations of individual stars from future extremely large telescopes, e.g.][]{Deep_et_al_11}.

\subsection{Effect of cluster formation model}
\label{sec:alt-phys}

\begin{figure*}
    \includegraphics[width=\textwidth]{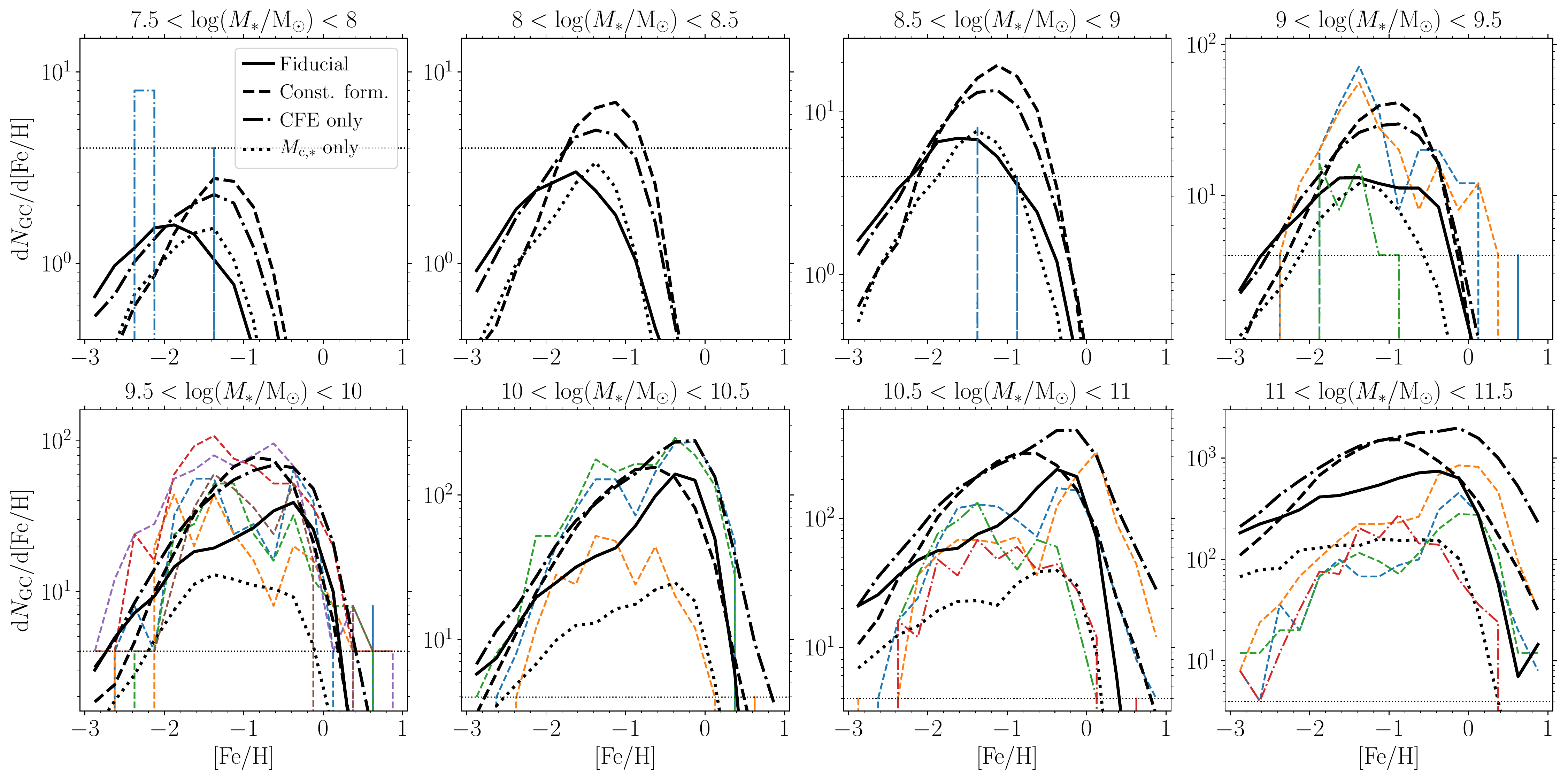}
    \caption{Average GC metallicity distributions for each of the E-MOSAICS GC formation models and GC ages $>8 \Gyr$. Panels show different galaxy mass ranges as in Fig.~\ref{fig:met_dists}. Solid black lines show the fiducial formation model (as in Fig.~\ref{fig:met_dists}), dashed black lines show the constant formation model (10 per cent CFE, power-law initial GC mass function), dash-dotted black lines show the CFE-only model (varying CFE, power-law initial GC mass function) and dotted black lines show the $\Mcstar$-only model (varying upper initial truncation mass, 10 per cent CFE). Coloured lines show observed distributions for comparison. In the figure we only show Local Group and Fornax cluster galaxies for clarity. In each panel, the dashed horizontal lines shows the limit of 1 GC per metallicity bin.}
    \label{fig:altphys}
\end{figure*}

\begin{figure*}
    \includegraphics[width=\textwidth]{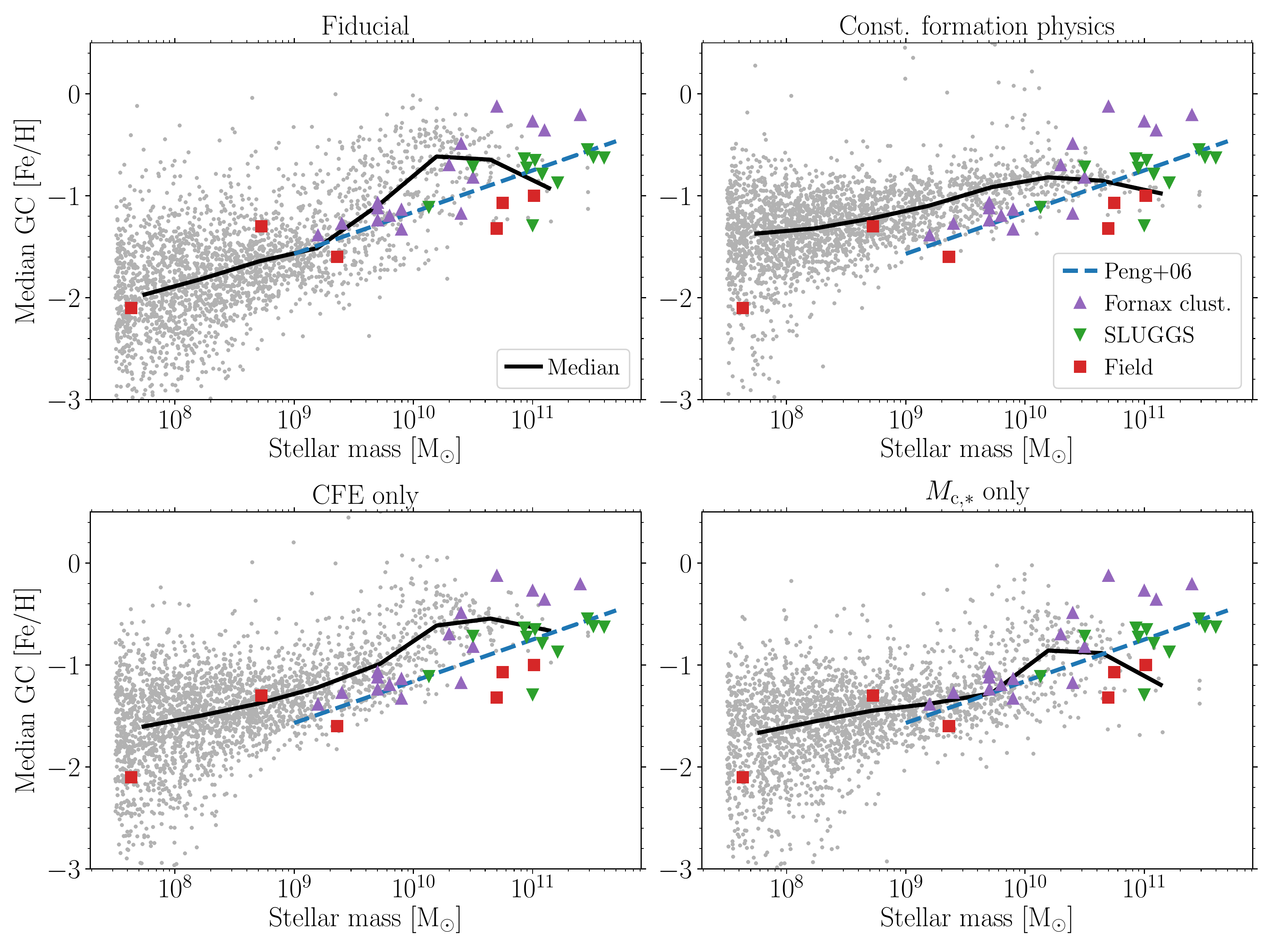}
    \caption{Median GC population metallicity as a function of galaxy mass for each GC formation model (Fig.~\ref{fig:altphys}). Individual E-MOSAICS galaxies are shown by the grey points, with the median shown by the solid black lines. Blue dashed lines show the relation for Virgo cluster galaxies from \citet{Peng_et_al_06}. Coloured points are the observed Local Group, Fornax cluster and SLUGGS survey galaxies from Fig.~\ref{fig:met_dists}. Except for the overabundance of metal-rich GCs in galaxies with $M_\ast \sim 10^{10.5} \Msun$ (see Sections~\ref{sec:emosaics} and \ref{sec:met_dists}), the median relation for the fiducial GC model (top left) shows reasonable agreement with the observed median metallicity relation. In contrast, the CFE-only model (bottom left) is offset to metallicities which are slightly too high ($\sim 0.3$~dex), while the constant formation (top right) and $\Mcstar$-only (bottom right) models show median relations which are too flat with galaxy mass. Appendix~\ref{app:rad_lims} shows the same figure but with a $20\kpc$ radius limit (as in Fig.~\ref{fig:Rcut}) to demonstrate the effect of the observational aperture on the median metallicity.}
    \label{fig:median_met}
\end{figure*}

A benefit of the E-MOSAICS approach to modelling galaxy and GC formation is that multiple star cluster formation models can be tested simultaneously in parallel within the same simulation.
In this section we test the effect of the different assumptions about the formation physics of GCs on the resulting GC metallicity distributions.
We consider four formation models (i.e.\ two settings for each of the CFE and $\Mcstar$ models):
varying CFE and $\Mcstar$ (fiducial model), constant CFE and power-law initial mass function (`constant formation' model), varying CFE and power-law initial mass function (`CFE-only' model) and constant CFE and varying $\Mcstar$ (`$\Mcstar$-only' model).

In Fig.~\ref{fig:altphys}, we compare the metallicity distributions for the four formation models as a function of galaxy mass (as in Fig.~\ref{fig:met_dists}).
As discussed by \citet{Bastian_et_al_20}, the number of GCs predicted varies significantly between the formation models, with CFE-only and constant formation models (i.e.\ variations without an upper mass truncation) predicting the most GCs and the $\Mcstar$-only model predicting the fewest.
The shapes of the GC metallicity distributions also vary between the different models.
At galaxy masses $M_\ast < 10^9 \Msun$, the three alternative formation models peak at higher metallicities than the fiducial model, while at $M_\ast \gtrsim 10^{10} \Msun$ the constant-formation model peaks at lower metallicities than the fiducial model.

We quantify this further in Fig.~\ref{fig:median_met}, by comparing the median GC metallicity as a function of galaxy mass with observed galaxies.
We find the fiducial model (top left panel) agrees well with the median GC metallicity in observed galaxies, other than at $M_\ast \sim 10^{10.5} \Msun$ where the simulated galaxies produce too many high-metallicity GCs.
At very large masses ($M_\ast > 10^{11} \Msun$) the simulated galaxies have median GC metallicities much lower than the Fornax cluster galaxies.
As discussed in Section~\ref{sec:met_dists}, this is related to limited ACS Fornax Cluster Survey field of view.
In Appendix~\ref{app:rad_lims} we show the predicted median metallicities when adopting a $20 \kpc$ radius limit. With this limit the median GC metallicities of massive galaxies are in good agreement with Fornax/Virgo clusters and SLUGGS survey galaxies, but over-predict the metallicities relative to field galaxies (Milky Way, M31, M81).

The comparison of the four different formation models in Fig.~\ref{fig:median_met} reveals how the CFE and $\Mcstar$ models affect the GC metallicity-galaxy mass relation.
The models with an environmentally-dependent CFE (fiducial, CFE-only) produce relations with slopes in reasonable agreement with the observed galaxies, while models with a constant CFE (constant, $\Mcstar$-only) produce relations with slopes that are too flat.
This is because star formation generally occurs at higher pressures in higher mass galaxies (at least at the peak of the star-formation rate), thus resulting in higher CFE in the environmentally-dependent model \citep[though this will vary significantly depending on galaxy formation history, for further discussion see][]{P18}.
Models without an upper GC mass function truncation (constant, CFE-only) generally have too high GC metallicities for low-mass galaxies ($M_\ast < 10^{10} \Msun$), while models with an environmentally-dependent $\Mcstar$ are in better agreement with observed galaxies at the same masses (i.e.\ the $\Mcstar$ model may suppress GC formation at high metallicities).
Thus, the CFE model influences the slope of the GC metallicity-galaxy mass relation, while the upper mass function truncation (or lack of) influences the normalisation.

Comparison of the GC formation models in Fig.~\ref{fig:altphys} also suggests that the underabundance of low metallicity GCs ($\FeH < -1$) in galaxies with masses $9.5 < \log (M_\ast / \mathrm{M}_{\sun}) < 10$ in the fiducial model (noted in Section~\ref{sec:met_dists}) could be due to a suppression of GC numbers by $\Mcstar$.
Comparing the fiducial and CFE-only models (i.e.\ with and without an $\Mcstar$ model), the CFE-only model appears in better agreement with the Fornax cluster galaxies.
However, we note that overall the upper truncation masses of GC systems in the fiducial model are in very good agreement with Virgo cluster galaxies, including this mass range \citep{Hughes_et_al_22}.
Additionally, the CFE-only model over-predicts low metallicity GC numbers in higher mass galaxies ($M_\ast > 10^{10.5} \Msun$).
In the future, detailed comparisons of the upper truncation mass of GC systems as a function of metallicity and galaxy mass between observed and simulated galaxies may help to resolve the issue.
The problem could also stem from the EAGLE model (i.e.\ too few low metallicity star particles, within which the GCs form) as galaxies in the `Recalibrated' EAGLE model have median stellar metallicities $\sim 0.2$-$0.3$~dex higher than observed within this mass range \citep[which itself could also be due to uncertainty in nucleosynthetic yields or absence of metal mixing between SPH particles]{S15}.
Alternatively, our GC luminosity selection (Section~\ref{sec:GC_selection}) may be too bright in comparison to the observed galaxies in this mass range.

\subsection{GC metallicity distribution bimodality}
\label{sec:bimodality}

\begin{figure*}
    \includegraphics[width=\textwidth]{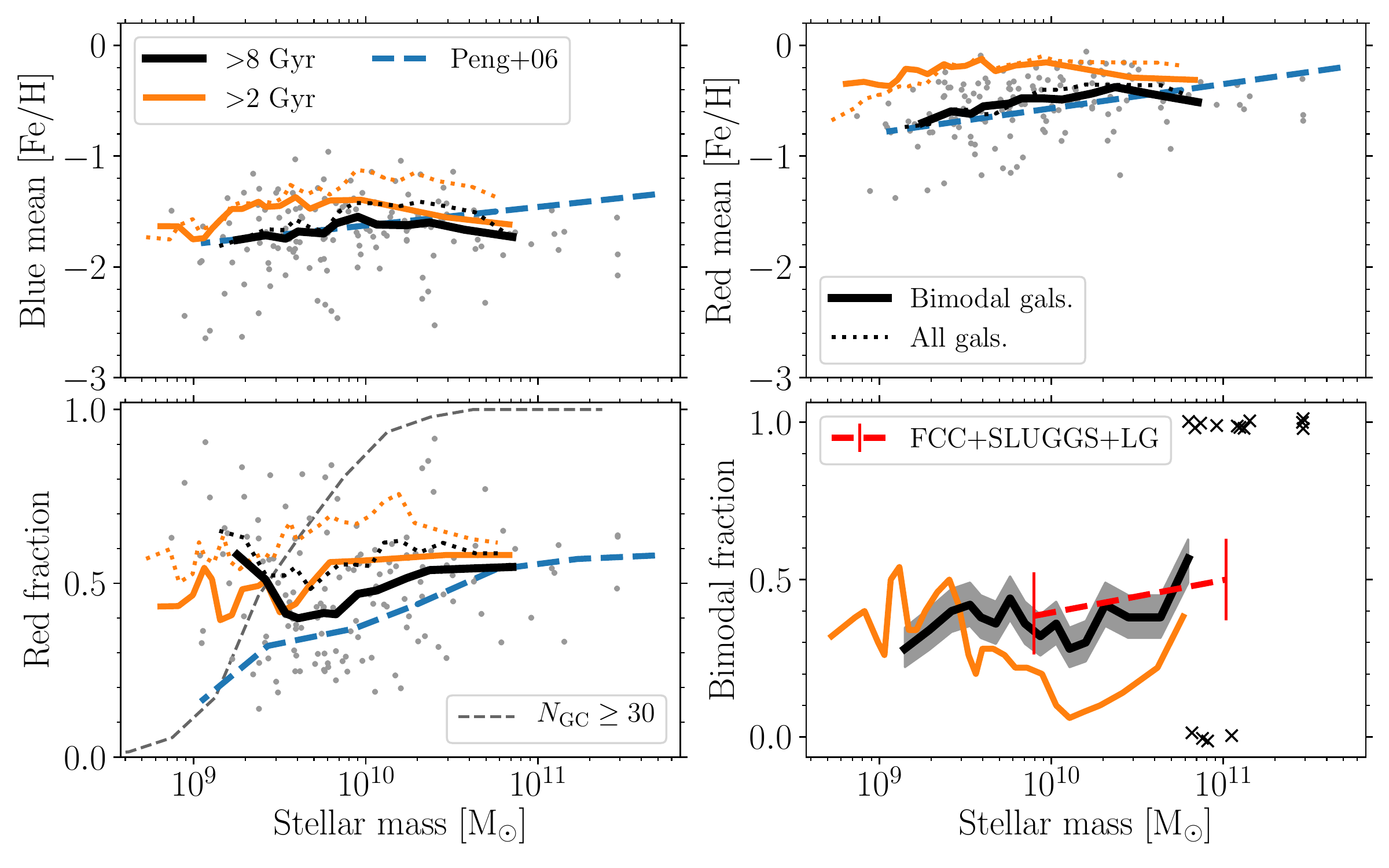}
    \caption{Results of applying Gaussian mixture modelling tests for bimodality \citep{Muratov_and_Gnedin_10} to the E-MOSAICS GC metallicity distributions (for galaxies with $\geq 30$ GCs). The top panels show the mean `blue' (top left) and `red' (top right) metallicity peaks as a function of galaxy mass, while the bottom left panel shows the fraction of GCs within the red peak as well as the fraction of all galaxies with $\geq 30$ GCs (dashed grey line; ages $>8 \Gyr$). Grey points show individual E-MOSAICS galaxies (GC ages $> 8 \Gyr$) which satisfy the GC bimodality tests.
Thick solid lines show the running medians for only bimodal galaxies, while dotted lines show results for all galaxies (though note the `red fraction' is not meaningful for unimodal distributions). Black and orange lines show GC age limits of $>8 \Gyr$ and $>2 \Gyr$, respectively.
Dashed blue lines show results from Virgo cluster galaxies \citep{Peng_et_al_06}. The bottom right panel shows the fraction of all galaxies (running medians) which satisfy the GC bimodality tests. The grey shaded region shows the uncertainty on the bimodal fraction for the $>8 \Gyr$ subsample and black crosses indicate the bimodality of the highest mass galaxies (above the median mass of the highest galaxy mass bin; i.e.\ bimodal$=1$, non-bimodal$= 0$). The observed bimodal fractions for Fornax cluster galaxies \citep{Fahrion_et_al_20b}, SLUGGS survey galaxies \citep{Usher_et_al_12} and Local Group galaxies (Milky Way, M31, M81) are shown by the red dashed line, with errorbars showing the uncertainties. Uncertainties on the bimodal fractions were calculated using binomial statistics.}
    \label{fig:bimodality}
\end{figure*}

We now test whether the E-MOSAICS GC formation model produces bimodal GC metallicity distributions in a way that is consistent with observations.
We use the Gaussian mixture modelling (GMM) algorithm from \citet{Muratov_and_Gnedin_10} to test whether a unimodal or bimodal distribution is preferred for the GC metallicity distribution of each galaxy with at least 30 GCs.
Following \citet{Muratov_and_Gnedin_10}, we consider a distribution to be bimodal if the bimodal solution is preferred with a probability $p > 0.9$, the distribution has a negative kurtosis and the two peaks are separated by a relative distance $D > 2$ (defined as $D = \| \mu_1 - \mu_2 \| / [ ({\sigma_1}^2 + {\sigma_2}^2)/2 ]^{1/2}$, where $\mu$ and $\sigma$ are the mean and standard deviation of a Gaussian distribution, respectively).

In Fig.~\ref{fig:bimodality} we compare the outcomes of the bimodality tests, showing the metallicities of the inferred `blue' (low metallicity, top left panel) and `red' (high metallicity, top right panel) peaks and `red' GC fraction for bimodal galaxies (bottom left panel), as well as the fraction of bimodal galaxies as a function of galaxy mass (bottom right panel).
We show the results for both GC age limits of $>8 \Gyr$ (standard sample, black lines) and $>2 \Gyr$ (including younger disc GCs, orange lines).
Comparing the GC age $>8 \Gyr$ samples (the most relevant comparison for cluster galaxies given they are generally old due to early star formation quenching, e.g.\ \citealt{Gallazzi_et_al_21}), we find very good agreement with the Virgo cluster results from \citet{Peng_et_al_06} for the trend of mean blue and red GC peak metallicities with galaxy mass (top panels), showing a trend of higher-metallicity peaks in higher-mass galaxies.
The drop or flattening in peak metallicities at high galaxy masses, relative to the Virgo cluster trend, is a result of the absence of a radius limit in the GC selection (see Section~\ref{sec:alt-phys} and Appendix~\ref{app:rad_lims}).
A similar flattening in the location of the peak metallicities was found for galaxies with $M_\ast \gtrsim 10^{11} \Msun$ (halo masses $\gtrsim 10^{13} \Msun$) by \citet{Choksi_Gnedin_and_Li_18}.
With an age limit of $>2 \Gyr$ the peak metallicities are $\approx 0.1$-$0.2$~dex higher, with the largest difference for the metal-rich peak in low-mass galaxies.
This is a result of galaxy downsizing, where lower-mass galaxies form more of their mass at later times \citep{Bower_et_al_92, Gallazzi_et_al_05}.

The bottom left panel of Fig.~\ref{fig:bimodality} shows the fraction of GCs in the red (high metallicity) peak for each bimodal galaxy.
For galaxy stellar masses $> 10^{9.5} \Msun$ and a $>8 \Gyr$ age limit we find good agreement with Virgo cluster galaxies \citep{Peng_et_al_06}, with a red fraction that increases with galaxy mass and plateauing at $\approx 0.55$.
With a $>2 \Gyr$ age limit the red fraction is only slightly higher than for a $>8 \Gyr$ limit.
At lower galaxy masses the red fraction appears to increase (for ages $>8 \Gyr$), unlike the observed fraction.
For a number of these low-mass galaxies the `red' peak occurs at $\FeH \sim -1.4$, which is similar to the blue peak in most other galaxies.
The occurrence of a second `blue' peak at lower metallicities in these galaxies could just be due to randomness where the number of GCs is low (i.e.\ $\sim 30$ GCs in total).
Indeed, for galaxies with a blue mean $\FeH \sim -2.5$ the typical number of GCs in the blue peak is only $\sim 10$ (with a range of $3$-$20$).
However, we also caution that at such masses ($\sim 10^9 \Msun$) we are comparing a biased sample of galaxies due to our requirement for a minimum number of GCs (30) to perform the bimodality tests.
The fraction of galaxies with at least 30 GCs is shown as a thin dashed line in the figure and is $<10$ per cent at $M_\ast = 10^9 \Msun$.
At similar masses, the median GC metallicity is $\FeH \approx -1.5$ (Fig.~\ref{fig:median_met}), thus the selection on GC numbers leads to a highly-biased sample of low-mass galaxies with a significant number of metal-rich GCs.
For completeness, the dotted lines in the figure show the results of the GMM tests for all galaxies, rather than only bimodal galaxies, though considering blue and red `peaks' for a unimodal distribution is not meaningful and simply reflects that the distribution is not a perfect Gaussian.
Comparing the GMM tests for all galaxies does not significantly change the peak metallicities, but does increase the red fractions to $>50$ per cent.

\begin{figure}
    \includegraphics[width=\columnwidth]{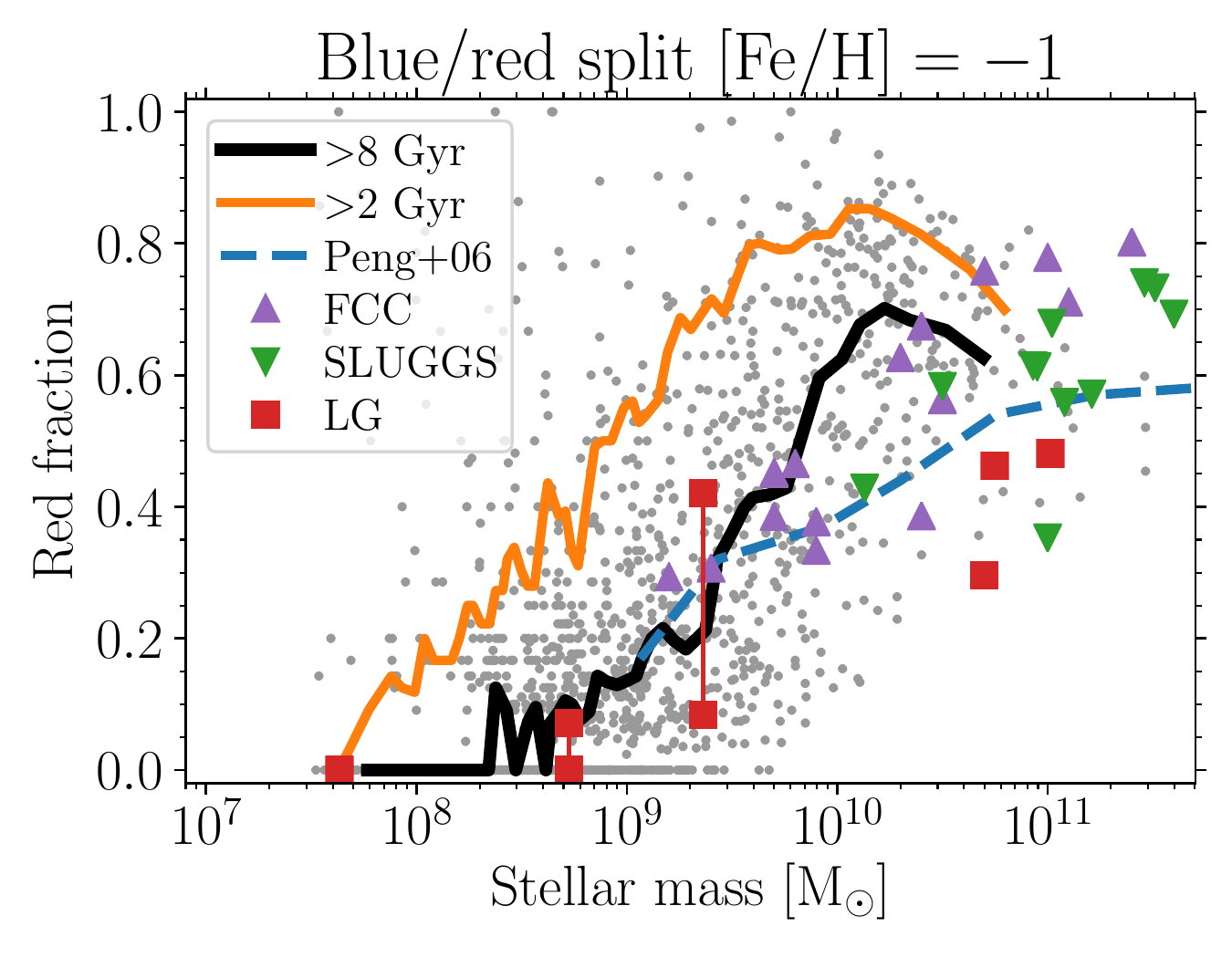}
    \caption{Fraction of `red' (high metallicity) GCs as a function of host galaxy stellar mass when adopting a fixed split between `red' and `blue' populations at $\FeH = -1$. Grey points show results for individual E-MOSAICS galaxies with at least 5 GCs (GC ages $>8 \Gyr$). This solid lines show running medians using age limits of $>8 \Gyr$ (solid black line) and $>2 \Gyr$ (solid orange line). Coloured points show results with the same fixed metallicity split for observed galaxies (as in Fig.~\ref{fig:met_dists}) from the Fornax cluster (purple triangles), SLUGGS survey (green upside-down triangles) and Local Group (red squares). For the LMC and SMC (where GC ages are available) we show results using both $>8$ and $>2 \Gyr$ age limits (red squares connected by vertical lines; red fractions are higher for lower age limits in both cases). The dashed blue line shows results from Virgo cluster galaxies \citep{Peng_et_al_06} for reference (as in Fig.~\ref{fig:bimodality}), though we note these results were not obtained with the same fixed metallicity split.}
    \label{fig:red_frac_fixed}
\end{figure}

As an alternative, more general, comparison, in Fig.~\ref{fig:red_frac_fixed} we compare the fraction of red (high metallicity) GCs using a fixed split between the populations at $\FeH = -1$ (a similar approach was used by \citealt{Choksi_and_Gnedin_19b} for comparing non-bimodal galaxies).
Though using fixed metallicity split ignores the increasing mean GC metallicity with galaxy mass, it can be applied to all galaxies, rather than only those with bimodal GC populations.
Similarly, it avoids issues where the `red' GC peak for some galaxies occurs at similar metallicities to typical blue GC peaks.
For a GC age limit $>8 \Gyr$ (grey points, solid black line), the simulations show good agreement with observed galaxies at galaxy stellar masses $\lesssim 10^{9.5} \Msun$ and $\gtrsim 10^{10.5} \Msun$.
At $M_\ast \approx 10^{10} \Msun$, simulated galaxies tend to have elevated red GC fractions by $\approx 0.2$, consistent with the elevated median GC metallicities at similar masses (Fig.~\ref{fig:median_met}).
With a GC age limit of $>2 \Gyr$ (solid orange line), we find elevated red GC fractions compared to using a $>8 \Gyr$ limit, with a typical difference in the fractions ranging from $0.1$ to $0.4$ depending on galaxy mass (peaking at $\approx 10^{9.5} \Msun$). A difference of $\approx 0.34$ is also found for the LMC, with a smaller difference ($\approx 0.07$) for the SMC (red squares connected by vertical lines).
The differences between the simulation trends in Fig.~\ref{fig:red_frac_fixed} and the bottom right panel of Fig.~\ref{fig:bimodality} (e.g. the increasing red fraction for low-mass galaxies in Fig.~\ref{fig:bimodality}) can be explained by galaxies with both low and high red fractions tending not to have bimodal GC populations.

In the bottom right panel of Fig.~\ref{fig:bimodality} we show the fraction of galaxies with bimodal GC metallicity distributions as a function of galaxy mass.
We find that $37 \pm 2$ per cent of simulated GC distributions are bimodal (for an age limit $> 8 \Gyr$) and the fraction is slightly higher ($45 \pm 7$ per cent) for higher mass galaxies ($M_\ast > 10^{10.5} \Msun$).
For an age limit $> 2 \Gyr$, the bimodal fraction is significantly lower ($\approx 10$ per cent) at masses $\sim 10^{10} \Msun$. This difference is largely caused by galaxies failing the kurtosis test (i.e.\ having a positive kurtosis) due to the increase in metal-rich GC numbers.
For comparison we also show the combined bimodal fractions from the Fornax cluster \citep{Fahrion_et_al_20b}, SLUGGS survey \citep{Usher_et_al_12} and Local Group galaxies (MW, M31, M81).
The total bimodal fraction for the observed galaxies is $44^{+10}_{-9}$ per cent ($12 / 27$ galaxies).
Overall, we find the predicted bimodal fractions from the simulations are in very good agreement with observed fractions.

\begin{figure}
    \includegraphics[width=\columnwidth]{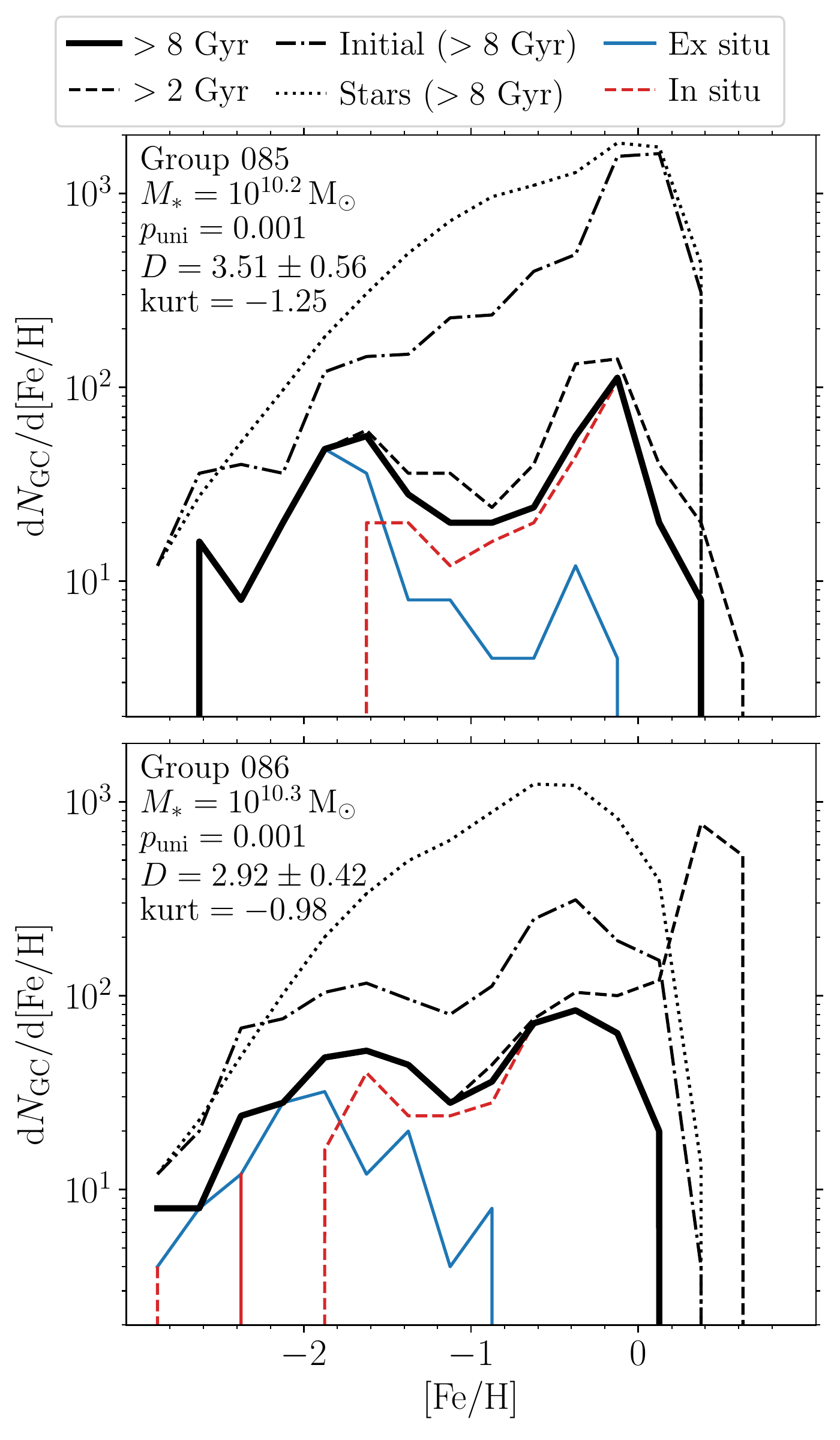}
    \caption{Two examples of bimodal GC metallicity distributions with different origins. The figure shows GC metallicity distributions for age limits of $>8 \Gyr$ (black solid line) and $>2 \Gyr$ (black dashed line), the initial distribution (ages $>8 \Gyr$, black dash-dotted line) and the distribution for star particles (ages $>8 \Gyr$, black dotted line, showing $\mathrm{d}M / \mathrm{d} \FeH$ with arbitrary normalisation). Solid blue and red lines show GCs (with ages $>8 \Gyr$) with an ex-situ and in-situ origin, respectively. For Group 85 (upper panel) bimodality results from cluster disruption, while for Group 86 (lower panel) bimodality is imprinted at cluster formation. Both galaxies are central galaxies of Milky Way-mass haloes ($M_{200} \approx 10^{12} \Msun$). Text in each panel shows the galaxy stellar mass and statistics from the GMM bimodality test \citep[unimodal distribution probability $p_\mathrm{uni}$, relative distance between Gaussian peak $D$ and kurtosis;][]{Muratov_and_Gnedin_10}.}
    \label{fig:bimodal_examp}
\end{figure}

\begin{figure}
    \includegraphics[width=\columnwidth]{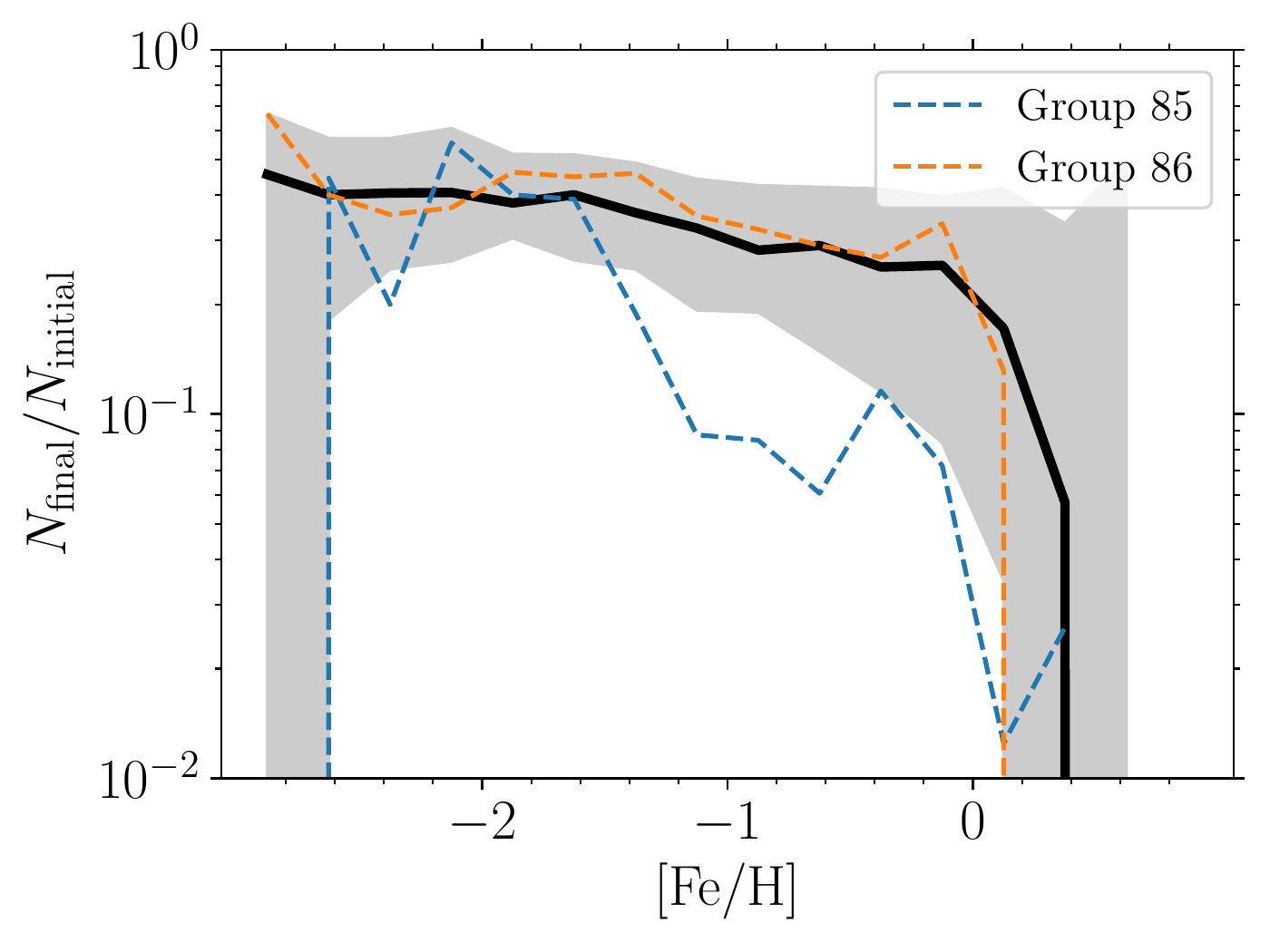}
    \caption{Ratio of final to initial number of GCs as a function of metallicity for Milky Way-mass haloes ($7 \times 10^{11} < M_\ast/\mathrm{M}_{\sun} < 3\times 10^{12}$, black line with grey shaded region showing 16\ts{th}-84\ts{th} percentile scatter between galaxies), highlighting the two galaxies from Fig.~\ref{fig:bimodal_examp} (dashed lines). The `initial' number of GCs accounts for stellar evolutionary mass loss (i.e.\ GCs that would pass the $z=0$ luminosity limit if there was no dynamical mass loss). While Group 86 closely follows the median relation, Group 85 has more effective cluster mass loss for $\FeH > -1.5$.}
    \label{fig:Nratios}
\end{figure}

To understand the origin of bimodal GC distributions in the simulated galaxies, we compare the metallicity distributions of two galaxies with clear bimodal distributions in Fig.~\ref{fig:bimodal_examp}. The galaxies were selected to be central galaxies in Milky Way-mass haloes ($M_{200} \approx 10^{12} \Msun$).
Despite relatively similar GC metallicity distributions for both galaxies, comparing the $z=0$ (solid line) and initial (dash-dotted line) distributions in each case shows a very different origin for bimodality.

In Group 86 (bottom panel) the bimodal GC distribution appears to be imprinted at cluster formation, with the initial distribution appearing as a scaled-up version of the $z=0$ distribution.
However, this is not reflected in the stellar metallicity distribution (for the same $>8 \Gyr$ age limit), which is unimodal and peaks at a similar metallicity to the `red' GC peak ($\FeH \approx -0.5$).
Interestingly, when also considering younger disc GCs (ages $> 2 \Gyr$) the GC metallicity distribution of this galaxy appears to be trimodal (with a third peak at $\FeH \approx 0.2$).
For this galaxy, the majority of the `blue' GC peak is in fact made up of in-situ clusters (unlike for Group 85), though there is a clear gradient in ex-situ fraction as a function of metallicity, which we will discuss further in Section~\ref{sec:bimod_origin}.

In Group 85 (top panel), neither the initial GC or stellar metallicity distributions are bimodal.
Instead, the GC distribution \textit{evolves} to be bimodal through dynamical mass loss.
We quantify this further in Fig.~\ref{fig:Nratios}, by comparing the ratio of the final ($z=0$) to initial number of GCs as a function of metallicity for central galaxies within $M_{200} \approx 10^{12} \Msun$ haloes.
For Group 86, GC mass loss is similar to the typical mass loss for such galaxies, being slightly more effective at high metallicities ($\FeH \sim 0$) than at lower metallicities ($\FeH \sim -2$; a result of the more disruptive, higher-density environment in which metal-rich GCs are formed; see \citealt{P18} and \citealt{K19a} for further discussion within the context of the E-MOSAICS simulations).
However, for Group 85 cluster mass loss is particularly effective at $\FeH \sim -1$ (i.e.\ at the trough in the $z=0$ distribution), which results in the evolution into a bimodal metallicity distribution.

\begin{figure*}
    \includegraphics[width=0.49\textwidth]{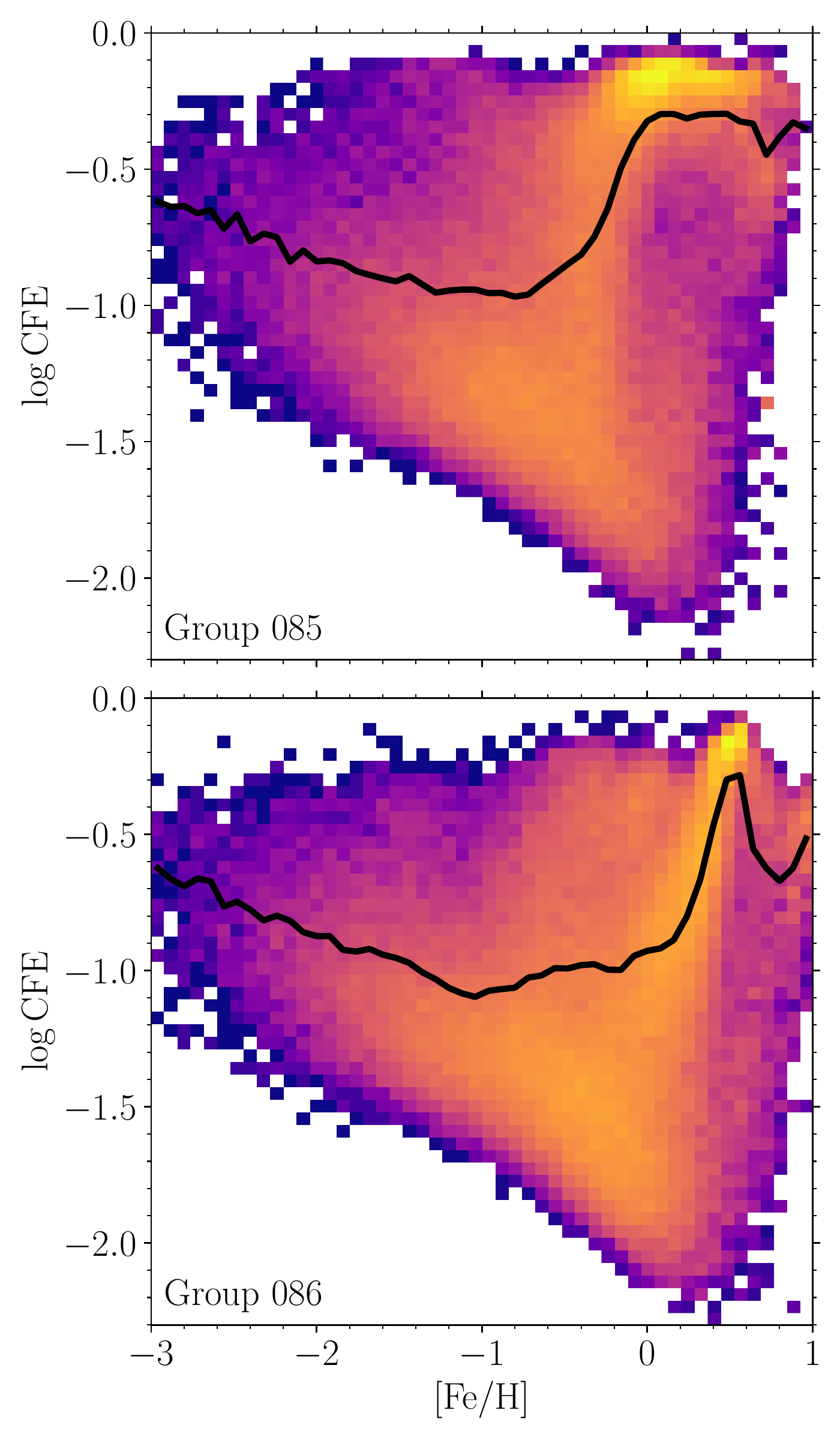}
    \includegraphics[width=0.49\textwidth]{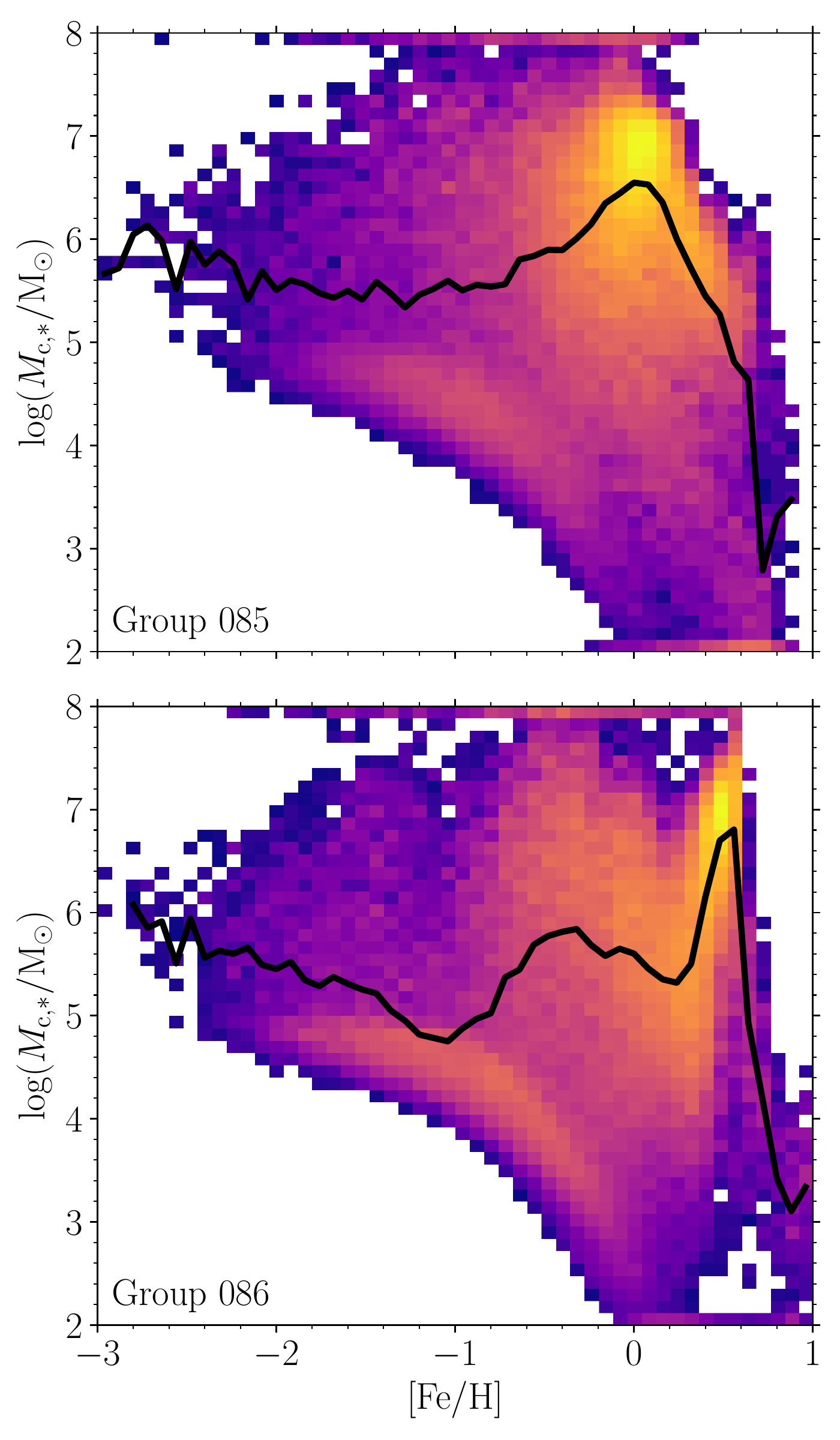}
    \caption{Distribution of CFE (left) and $\Mcstar$ (right) as a function of metallicity for all star particles with ages $>8 \Gyr$ that reside in the central galaxy of Groups 85 and 86 (i.e.\ galaxies in Fig.~\ref{fig:bimodal_examp}; top and bottom panels, respectively). Colours in the 2D histograms show logarithmic scales for number of star particles. For $\Mcstar$, the histograms are weighted by the CFE of the star particles. Particles with $\Mcstar < 10^2$ and $> 10^8 \Msun$ are shown at $10^2$ and $10^8 \Msun$ in the histogram, respectively. Solid lines show the average CFE and median CFE-weighted $\Mcstar$ as a function of metallicity.}
    \label{fig:form_props}
\end{figure*}

We investigate the origin of the bimodal initial GC distribution for Group 86, and lack of one in the initial distribution for Group 85, in Fig.~\ref{fig:form_props} by comparing the distributions of the cluster formation properties CFE (left panel) and $\Mcstar$ (right panel) as a function of metallicity.
Both galaxies show a decline in CFE from $\FeH = -3$ to $-1$, meaning GCs are least likely to form at $\FeH \approx -1$.
This decline is driven by the decreasing density threshold for star formation \citep[see figure 3 in][]{P18} which models the transition from a warm to cold interstellar medium that is expected to occur at lower densities for higher metallicities \citep{Schaye_04}, meaning at very low metallicities the fraction of star formation in clusters (CFE) is high.
The CFE in both galaxies then experiences an increase at higher metallicities as a result of high natal gas pressures at small galactic radii \citep[which also tends to result in GCs with elevated $\alpha$-abundances,][]{Hughes_et_al_20}.
However, the initially bimodal GC distribution in Group 86 is not completely caused by a change in the CFE (at least in this case), as the typical CFE ($\approx 10$ per cent) is relatively similar across the range $-1.5 < \FeH < -0.5$ with only a small decrease at $\FeH \approx -1.1$.
Instead, it appears to be further driven by an evolution in $\Mcstar$ (right panel in the figure, which itself also depends on the CFE), such that fewer GCs are formed at high enough masses that they pass the luminosity selection at $z=0$.
We find that the typical $\Mcstar$ for Group 86 falls below $10^5 \Msun$ at $\FeH \approx -1.1$, i.e.\ at the same metallicity where the trough in the initial (and indeed final) GC metallicity distribution is found for this galaxy.
This is not the case for Group 85, where the typical $\Mcstar$ remains relatively similar ($\approx 10^{5.5} \Msun$) at metallicities $\FeH < -0.5$.
Therefore, for some galaxies, variations in cluster formation properties (CFE and $\Mcstar$) as a function of metallicity can result in initially-bimodal GC metallicity distributions.

\section{Discussion}
\label{sec:discussion}

\subsection{Origin of bimodal GC metallicity distributions}
\label{sec:bimod_origin}

\begin{figure}
    \includegraphics[width=\columnwidth]{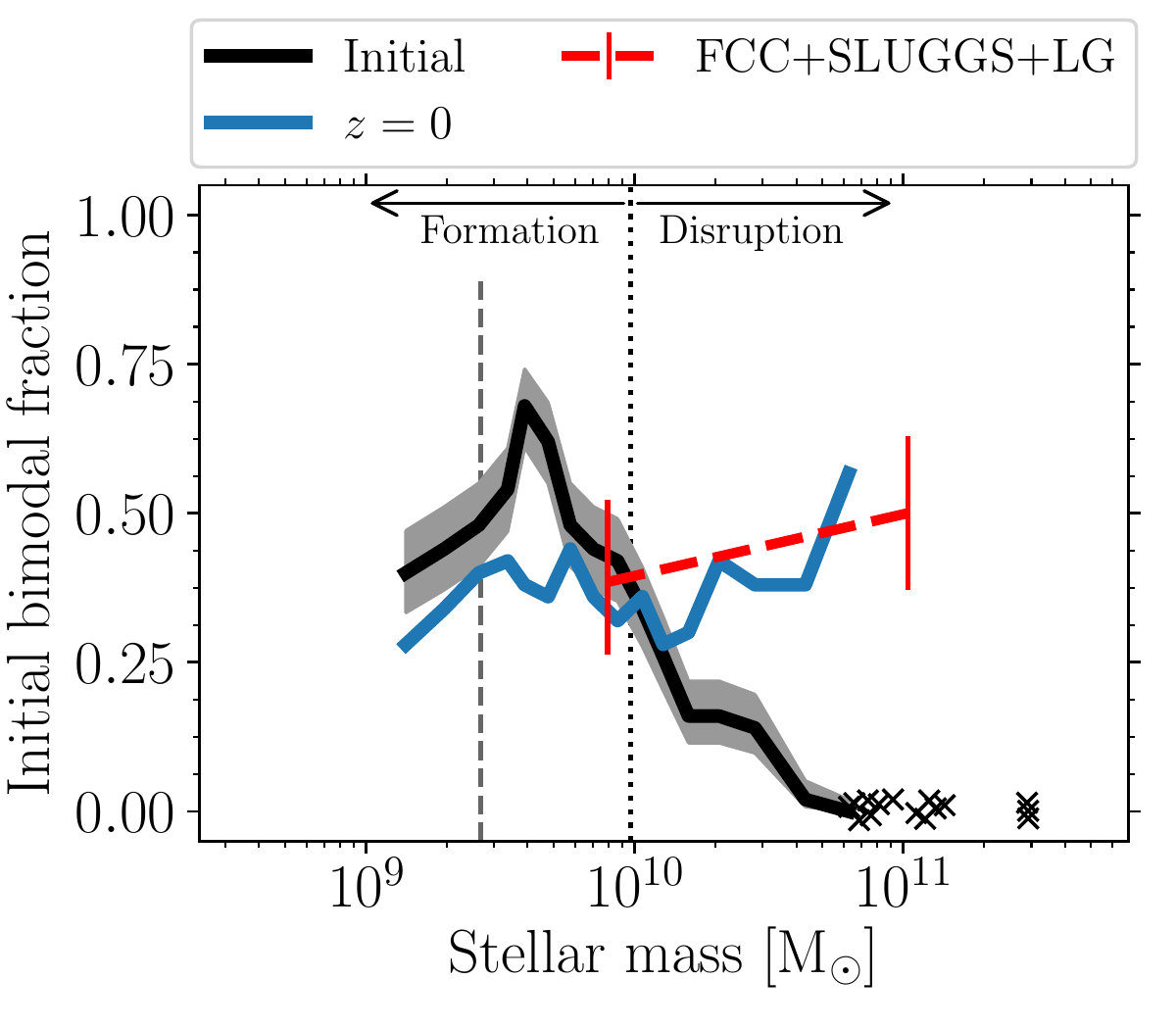}
    \caption{Fraction of simulated galaxies with bimodal GC metallicity distributions, as in the bottom right panel of Fig.~\ref{fig:bimodality}, but instead showing the results of the bimodality test for the initial GC metallicity distributions of those galaxies (i.e.\ only galaxies with $N_\mathrm{GC} \geq 30$ at $z=0$; black line, with grey shaded region showing the binomial confidence interval).
The solid blue line shows the $z=0$ bimodal fractions from Fig.~\ref{fig:bimodality}.
The red dashed line shows observed $z=0$ fractions as in Fig.~\ref{fig:bimodality}.
The vertical dotted line approximately indicates the mass at which the origin of GC bimodality in the simulations transitions from GC formation (lower masses) to GC disruption (higher masses).
The vertical dashed grey line shows the mass below which less than 50 per cent of galaxies satisfy the GC number limit (see bottom left panel of Fig.~\ref{fig:bimodality}).
}
    \label{fig:bimodality_init}
\end{figure}

Most previous works have focussed on GC formation as the origin of bimodal metallicity distributions or otherwise do not separately consider the effects of cluster formation and disruption.
In general, bimodality has been suggested to occur through discrete episodes of GC formation, e.g.\ via galaxy mergers \citep[e.g.][]{Zepf_and_Ashman_93, Muratov_and_Gnedin_10} or truncating the formation of metal-poor GCs below some redshift \citep[e.g.][]{Forbes_Brodie_and_Grillmair_97, Beasley_et_al_02, Strader_et_al_05, Griffen_et_al_10}.
Models where GC disruption is environmentally independent have nevertheless been successful in reproducing the observed peak metallicities of red and blue GC populations \citep{Choksi_Gnedin_and_Li_18}.

In the E-MOSAICS model GCs may form in a galaxy where star formation is sufficiently intense, and thus there is no explicit truncation in GC formation times (other than that which occurs `naturally' during the formation of galaxies, such as star formation quenched by stellar/AGN feedback or gas exhaustion, which will therefore vary significantly from galaxy to galaxy depending on their formation histories).
As shown in Section~\ref{sec:bimodality}, in the E-MOSAICS simulations GC metallicity bimodality arises in different ways: through variations in cluster formation (efficiency and upper mass truncation) as a function of metallicity creating an initially bimodal distribution, or cluster disruption creating a bimodal distribution from an initially unimodal distribution.

To determine whether cluster formation or disruption is more important in the origin of bimodal metallicity distributions in the E-MOSAICS model, in Fig.~\ref{fig:bimodality_init} we show the results of performing GMM tests on the initial GC metallicity distributions for galaxies in Fig.~\ref{fig:bimodality}.
Unlike for the $z=0$ bimodal fractions which are nearly constant at all galaxy masses (Fig.~\ref{fig:bimodality}), the initial bimodal fractions strongly depend on galaxy mass.
For $M_\ast \gtrsim 10^{9.5} \Msun$ (where $>50$ per cent of the galaxy population is sampled) the initial bimodal fractions decline with increasing galaxy mass ($\approx 60$ to $0$ per cent from $10^{9.5}$ to $10^{10.5} \Msun$).
In fact for galaxies with stellar masses $M_\ast > 10^{10.5} \Msun$, no galaxies have initially bimodal GC metallicity distributions, despite $\approx 40$ per cent of such systems being bimodal at $z=0$.

For $M_\ast \gtrsim 10^{10} \Msun$ (where the initial bimodal fraction falls below the $z=0$ fraction), Fig.~\ref{fig:bimodality_init} shows that cluster mass loss is therefore the most important factor for determining whether a GC metallicity distribution will become bimodal in the E-MOSAICS model.
A similar point was made by \citet{Kruijssen_15} who argued that GC destruction, which sets the specific frequency-metallicity relation (Section~\ref{sec:TN-met}), is responsible for observed GC metallicity distributions.
However, for galaxies with masses $M_\ast < 10^{10} \Msun$ the inverse appears to be the case:
More galaxies (47 per cent) have initially bimodal GC distributions than at $z=0$ (37 per cent).
Thus in low-mass galaxies GC evolution decreases the fraction of bimodal galaxies.

In principle, determining whether formation or disruption resulted in the GC metallicity distribution of observed galaxies could be tested by surveying the field stars from destroyed GCs.
In fact, the multiple stellar populations observed in Galactic GCs \citep[light-element abundance spreads that appear to be unique to GCs, see reviews by][]{Gratton_Carretta_and_Bragaglia_12, Charbonnel_16, Bastian_and_Lardo_18} potentially provide such a method.
By searching for their unique chemical fingerprints, former GC stars have been identified in the stellar halo \citep{Martell_and_Grebel_10, Martell_et_al_16, Koch_Grebel_and_Martell_19, Horta_et_al_21b} and bulge \citep{Schiavon_et_al_17} of the Milky Way.

Interestingly, in the inner Galaxy the nitrogen-rich (potentially former GC) stars peak at a metallicity $\FeH \approx -1$ \citep{Schiavon_et_al_17}, unlike the Galactic GC system which has a trough at a similar metallicity \citep[see figure 9 in][]{Schiavon_et_al_17}.
Thus the N-rich stars do not appear to originate from \textit{surviving} GCs, but instead could be the remnants of destroyed GCs.
This suggests that the bimodal metallicity distribution of Milky Way's GC system may be a result of GC disruption, rather than formation, in line with the expectations for massive galaxies in the E-MOSAICS model.

\begin{figure}
    \includegraphics[width=\columnwidth]{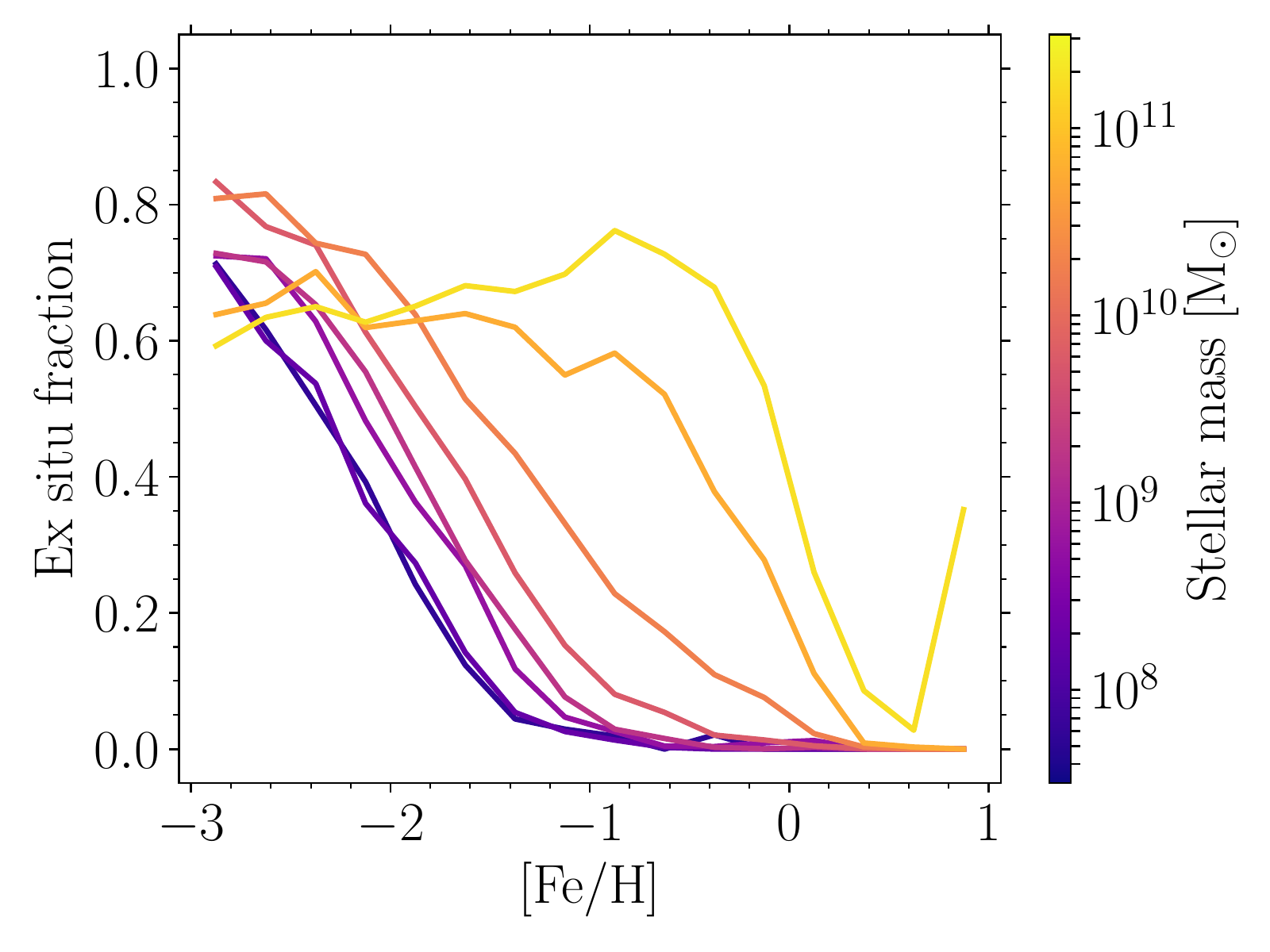}
    \caption{Average ex-situ GC fractions at $z=0$ as a function of metallicity and galaxy stellar mass. The galaxy stellar mass ranges are as in the panels of Fig.~\ref{fig:met_dists} (i.e.\ in $0.5$~dex ranges from $10^{7.5}$ to $10^{11.5} \Msun$) with line colours as indicated in the colourbar. The typical metallicity below which ex-situ GCs dominate (contribute $>50$ per cent) the population increases with galaxy mass, from $\FeH \approx -2.3$ for $M_\ast < 10^{8.5} \Msun$ to $\FeH \approx 0$ for $M_\ast > 10^{11} \Msun$.}
    \label{fig:exsitu}
\end{figure}

\begin{figure}
    \includegraphics[width=\columnwidth]{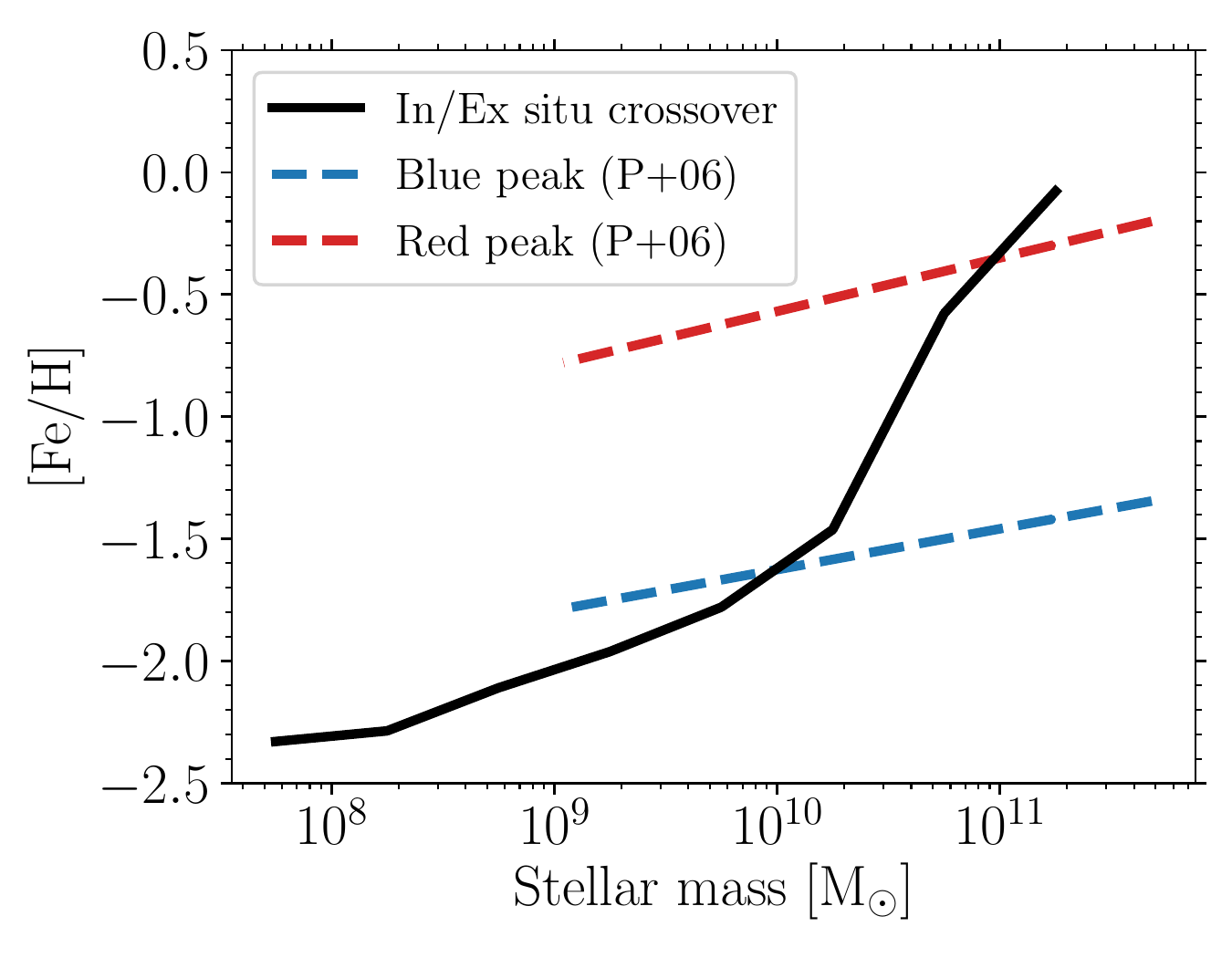}
    \caption{Metallicity below which ex-situ GCs become dominant ($>50$ per cent of population) as a function of galaxy mass (solid black line). The crossover metallicities were interpolated from the ex-situ fractions in Fig.~\ref{fig:exsitu}. Dashed blue and red lines show the blue and red peak metallicities, respectively, from \citet{Peng_et_al_06}. In general, the blue and red peak metallicities are unrelated to ex- and in-situ GC formation.}
    \label{fig:exsitu_cross}
\end{figure}

In the context of hierarchical galaxy formation, it has been suggested that metal-rich and metal-poor GCs represent in-situ and ex-situ formation \citep[e.g.][]{Cote_Marzke_and_West_98, Hilker_Infante_and_Richtler_99, Forbes_and_Remus_18}.
We test this idea in Fig.~\ref{fig:exsitu} by comparing the average ex-situ GC fractions as a function of metallicity and galaxy mass.
Star particles (and their associated GCs) are defined as in- or ex-situ based on the subhalo the parent gas particle was bound to in the snapshot immediately prior to star formation (i.e.\ the time when the gas particle was converted into a star particle).
Particles formed in subhaloes within the main progenitor branch are defined as in-situ formation.

The metallicity below which ex-situ GCs become dominant (fraction $>50$ per cent) shows a strong correlation with galaxy mass, from $\FeH \approx -2.3$ in low-mass galaxies ($M_\ast < 10^{8.5} \Msun$) to $\FeH \approx 0$ in massive galaxies ($M_\ast > 10^{11} \Msun$).
We show this directly in Fig.~\ref{fig:exsitu_cross}.
This is a consequence of the increasing contribution of mergers to the growth of galaxies at higher masses \citep[e.g.][]{Qu_et_al_17, Clauwens_et_al_18, Pillepich_et_al_18, Tacchella_et_al_19, Davison_et_al_20}.
As shown by \citet{Choksi_and_Gnedin_19b}, the fraction of ex-situ GCs reasonably follows the increasing ex-situ field star fraction with increasing galaxy mass.
Thus in massive galaxies ($M_\ast > 10^{11} \Msun$) GC bimodality is not directly related to in/ex-situ GC formation as the majority of GCs are formed ex-situ, even in the red peak.
Rather, the GC metallicities are related to the time of formation and the galaxy mass-metallicity relation \citep{Muratov_and_Gnedin_10, Kruijssen_15, Kruijssen_19}.
For galaxies with masses $\lesssim 5 \times 10^9 \Msun$, ex-situ GCs only become dominant at metallicities below the typical blue peak metallicity.
Only in galaxies with stellar masses of $\approx 10^{10.5} \Msun$ will the blue and red peaks (generally) correspond to ex-situ and in-situ GC formation.
Naturally, this will also vary from galaxy to galaxy, as shown in Fig.~\ref{fig:bimodal_examp} where ex-situ GCs dominate the blue peak for Group 85, but in-situ GCs dominate the blue peak for Group 86.
A correlation between the accreted fractions of metal-poor/-rich GCs and galaxy mass was similarly found by \citet{Choksi_and_Gnedin_19b}, despite the very different GC models, and thus appears to be a general feature of cosmological GC formation models based on `young star cluster' formation (see next section).

\subsection{GC metallicity distributions and the origin of GCs}
\label{sec:GC_origin}

The metallicity distributions of GC systems, bimodal distributions in particular, have been seen as critical in understanding the origin of GCs, with a number of different scenarios proposed to explain their origin.
Some works have investigated a direct connection between GC formation and gas-rich galaxy mergers \citep{Ashman_and_Zepf_92, Zepf_and_Ashman_93, Beasley_et_al_02, Muratov_and_Gnedin_10}, inspired by the the young star clusters with GC-like properties observed in merging and post-merger galaxies \citep[e.g.][]{Holtzman_et_al_92, Whitmore_et_al_93, Whitmore_and_Schweizer_95}.
In this scenario, many mergers of low-mass galaxies and few mergers of high-mass galaxies thus create the metal-poor and metal-rich GC distribution peaks, respectively.
Such models can reproduce many scaling relations of GC populations \citep{Li_and_Gnedin_14, Choksi_Gnedin_and_Li_18}.
However, young massive star clusters have been observed forming in a variety of environments (e.g.\ from regular spiral galaxies to star-burst dwarf galaxies), not just merging and interacting galaxies \citep{OConnell_Gallagher_and_Hunter_94, Larsen_and_Richtler_99, Hunter_et_al_00, Billett_Hunter_and_Elmegreen_02, Mora_et_al_09, Chandar_et_al_10, Annibali_et_al_11}.
Though galaxy mergers can help to drive conditions favourable for star cluster formation \citep[higher gas pressures and star formation rates, e.g.][]{Bekki_et_al_02, Bournaud_Duc_and_Emsellem_08, Lahen_et_al_20, Li_et_al_22}, massive star cluster formation may still occur in the absence of galaxy mergers.
Mergers may also promote the survival of star clusters by ejecting them from the disruptive environment of star-forming discs \citep{Kravtsov_and_Gnedin_05, Kruijssen_15}, and in this way merger-based models may have GC survival included in the formation model.
Updated versions of the \citet{Muratov_and_Gnedin_10} semi-analytic model (which model galaxies and GC populations by applying analytic scaling relations to dark matter-only cosmological simulations) instead consider GC formation during high halo accretion rates, rather than just during mergers \citep{Choksi_Gnedin_and_Li_18, Choksi_and_Gnedin_19b}.

Other works have suggested separate formation mechanisms for metal-poor and metal-rich GCs.
While metal-rich GCs may form in mergers or otherwise with the bulk of the field stars, metal-poor GCs have been suggested to form in low-mass dark matter haloes in the early Universe \citep{Peebles_84, Rosenblatt_Faber_and_Blumenthal_88, Ricotti_Parry_and_Gnedin_16}, whose formation is possibly truncated \citep[e.g.][]{Strader_et_al_05, Moore_et_al_06, Griffen_et_al_10} or caused \citep{Cen_01} by reionization.
As already discussed in previous works, the scenario has a number of fundamental issues.
Isolated GCs, residing in dark matter haloes yet to be accreted into a larger system, have not been observed \citep{di_Tullio_Zinn_and_Zinn_15, Mackey_Beasley_and_Leaman_16}.
Some GCs (i.e.\ those which always resided in weak tidal fields) might be expected to retain their dark matter halo, however GCs at large distances from their host galaxy are consistent having no dark matter halo \citep{Baumgardt_et_al_09, Lane_et_al_10, Conroy_Loeb_and_Spergel_11}.
Critically, to explain GC metallicities such GCs must self-enrich through supernovae feedback, which is inconsistent with the very small iron spreads \citep[$<0.1$~dex,][]{Carretta_et_al_09} in most\footnote{The very few Milky Way GCs which do contain large heavy element spreads are consistent with being the former nuclear clusters of accreted dwarf galaxies or rare cases of cluster-cluster mergers \citep[see][and references therein]{Pfeffer_et_al_21}.} GCs \citep{Peebles_84, Rosenblatt_Faber_and_Blumenthal_88}.

In this work, we show that a GC formation and evolution model which reproduces the observed scaling relations of young star cluster populations \citep[using the fiducial E-MOSAICS model]{Pfeffer_et_al_19b} also reproduces many properties of GC metallicity distributions when applied to galaxy formation in a cosmological context.
The model does not rely on separate formation mechanisms or truncated formation epochs for metal-poor and metal-rich GCs, nor preferential formation in galaxy mergers \citep[which generally are not a major contribution to GC formation in the E-MOSAICS model,][]{Keller_et_al_20}.
Instead, variations in GC metallicity distributions between galaxies may arise due to local environmental variations in GC formation properties and their subsequent evolution over time.

A `young star cluster' scenario for the origin of GCs can also explain many other properties of GC populations in galaxies, including the masses of GCs \citep{Kravtsov_and_Gnedin_05, Kruijssen_15, Li_et_al_17, Ma_et_al_20}, the fraction of stars contained in GCs \citep{Choksi_Gnedin_and_Li_18, Bastian_et_al_20}, the typical old ages of GCs \citep{Reina-Campos_et_al_19}, the age-metallicity relations of GC systems \citep{K19a, Li_and_Gnedin_19, Horta_et_al_21a}, the radial distributions of GC systems \citep{Reina-Campos_et_al_22a}, the high-mass truncation of GC mass functions \citep{Hughes_et_al_22}, the redder colours of more luminous GCs \citep[`blue tilt',][]{Usher_et_al_18, Choksi_and_Gnedin_19a, Kruijssen_19} and the UV luminosity function of proto-GCs at high redshifts \citep{Bouwens_et_al_21}.
The uncertainties that remain in this scenario are generally related to the difficulties in modelling both the small scales of GCs ($\sim$parsecs) and the large scales of galaxy populations ($\sim$megaparsecs) or particular numerical implementations \citep[e.g.\ such as the overabundance of low-mass GCs in the E-MOSAICS simulations,][]{P18}.
Considered together, this body of work suggests a common origin for GCs and young star clusters.

\section{Summary}
\label{sec:summary}

In this work, we investigate the GC metallicity distributions predicted by the E-MOSAICS simulations.
The main aim of the work is to test whether a common formation scenario for young star clusters and GCs can explain the observed GC metallicity distributions.
We find that:
\begin{enumerate}

\item The predicted GC metallicity distributions from the fiducial E-MOSAICS model generally agree well with the observed distributions as a function of galaxy mass (Section~\ref{sec:met_dists}).
In particular, the distributions for high-mass galaxies ($M_\ast > 10^{11} \Msun$) are in very good agreement with observed distributions once the field of view is accounted for (Fig.~\ref{fig:Rcut}).
However, simulated Milky Way-mass galaxies ($10.5 < \log(M_\ast / \rmn{M}_{\sun}) < 11$) often have too many metal-rich GCs compared to observed galaxies \citep[as already discussed in detail in][]{K19a}, which is suggested to be due to inadequate disruption of GCs through lack of substructure in galactic discs (see Section~\ref{sec:emosaics} for a brief summary).
The number of extremely low metallicity GCs ($\FeH \lesssim -2.5$) are also overpredicted (by a factor $\approx 3$), which may be a result of the resolution limits of the simulation.

\item The predicted relationship between GC specific frequency (number of GCs per unit mass) and metallicity shows a slope that is consistent with observed trends (Section~\ref{sec:TN-met}). In the simulations, this trend emerges due to an increased CFE at low metallicities and preferential disruption of high metallicity GCs in higher mass galaxies.
However, the under-disruption of high-metallicity GCs at small galactocentric radii is also evident in the specific frequency-metallicity relations.

\item Comparing the variations of the E-MOSAICS GC formation model (Section~\ref{sec:alt-phys}), we find that models with a constant CFE fail to reproduce the correlation between median GC system metallicity and galaxy mass, instead predicting a relation that is too flat, while a model with an environmentally-varying CFE but without an upper mass function truncation predicts median GC metallicities that are too high. Of the four formation models, only the fiducial E-MOSAICS model (with both an environmentally-varying CFE and $\Mcstar$) reasonably matches the slope and normalisation observed relation, except for the excess of high metallicity galaxies in galaxies with $M_\ast \sim 10^{10.5} \Msun$ (which appears in all models).

\item We use Gaussian mixture modelling \citep{Muratov_and_Gnedin_10} to test the predicted fraction of galaxies with bimodal distributions (Section~\ref{sec:bimodality}). We find that $37 \pm 2$ per cent of simulated GC distributions are bimodal, compared with $44^{+10}_{-9}$ per cent for observed galaxies. The mean metallicities of the metal-poor and metal-rich peaks, as well as the fraction of metal-rich GCs (for $M_\ast > 10^{9.5} \Msun$), are also in good agreement with the observed relations.

\item We demonstrate that bimodal GC metallicity distributions in E-MOSAICS galaxies are created in two ways (Section~\ref{sec:bimodality}): variations in cluster formation properties as a function of metallicity that form a bimodal distribution, and cluster disruption generating a bimodal distribution from an initially-unimodal distribution.
For high-mass galaxies in the simulation ($M_\ast > 10^{10} \Msun$), most initial GC metallicity distributions are unimodal, and thus bimodality occurs largely as a result of cluster disruption.
Based on the metallicity distribution of stars in the Milky Way which may originate from disrupted GCs \citep{Schiavon_et_al_17}, we suggest that the bimodal GC metallicity distribution of the Milky Way is similarly a result of cluster disruption, rather than formation (Section~\ref{sec:bimod_origin}).

\end{enumerate}

Overall, the fiducial E-MOSAICS model reproduces many features of observed GC metallicity distributions, with a few identified shortcomings.
Addressing the issues with insufficient GC disruption and overabundance of metal-rich GCs (for galaxies with $M_\ast \sim 10^{10.5} \Msun$) requires progress on two fronts: higher resolution and more detailed galaxy formation models (particularly with regard to modelling the cold interstellar medium, e.g. \citealt{Reina-Campos_et_al_22c}); and more realistic subgrid models for the dynamical evolution of GCs (currently the size evolution of GCs is not modelled, which is important for understanding how GCs respond to tidal shocks, e.g. \citealt{Gieles_et_al_06c, Gieles_and_Renaud_16, Webb_et_al_19, Martinez-Medina_et_al_22}).

We have shown that a scenario in which old GCs and young star clusters form and evolve via the same mechanisms is able to explain the metallicity distributions observed for GCs in nearby galaxies.
In view of this result, we conclude that different formation mechanisms or truncated formation epochs are not required to separately explain metal-poor and metal-rich GCs (Section~\ref{sec:GC_origin}).
Instead, GCs can be considered as the evolved, surviving analogues of young star clusters observed forming in galaxies at the present day.

\section*{Acknowledgements}

We thank Chris Usher for helpful discussions and providing SLUGGS survey data, Katja Fahrion for providing Fornax cluster data and the referee, Prof. Oleg Gnedin, for helpful comments.
JP is supported by the Australian government through the Australian Research Council's Discovery Projects funding scheme (DP200102574).
JMDK gratefully acknowledges funding from the German Research Foundation (DFG - Emmy Noether Research Group KR4801/1-1). JMDK and STG gratefully acknowledge funding from the European Research Council (ERC-StG-714907, MUSTANG).
COOL Research DAO is a Decentralised Autonomous Organisation supporting research in astrophysics aimed at uncovering our cosmic origins.
RAC is supported by the Royal Society.
This work used the DiRAC Data Centric system at Durham University, operated by the Institute for Computational Cosmology on behalf of the STFC DiRAC HPC Facility (\url{www.dirac.ac.uk}). This equipment was funded by BIS National E-infrastructure capital grant ST/K00042X/1, STFC capital grants ST/H008519/1 and ST/K00087X/1, STFC DiRAC Operations grant ST/K003267/1 and Durham University. DiRAC is part of the National E-Infrastructure.
The work also made use of high performance computing facilities at Liverpool John Moores University, partly funded by the Royal Society and LJMU's Faculty of Engineering and Technology.

\section*{Data Availability}

The data underlying this article will be shared on reasonable request to the corresponding author.



\bibliographystyle{mnras}
\bibliography{emosaics} 



\appendix

\section{Median metallicities with radius limits}
\label{app:rad_lims}

\begin{figure*}
    \includegraphics[width=\textwidth]{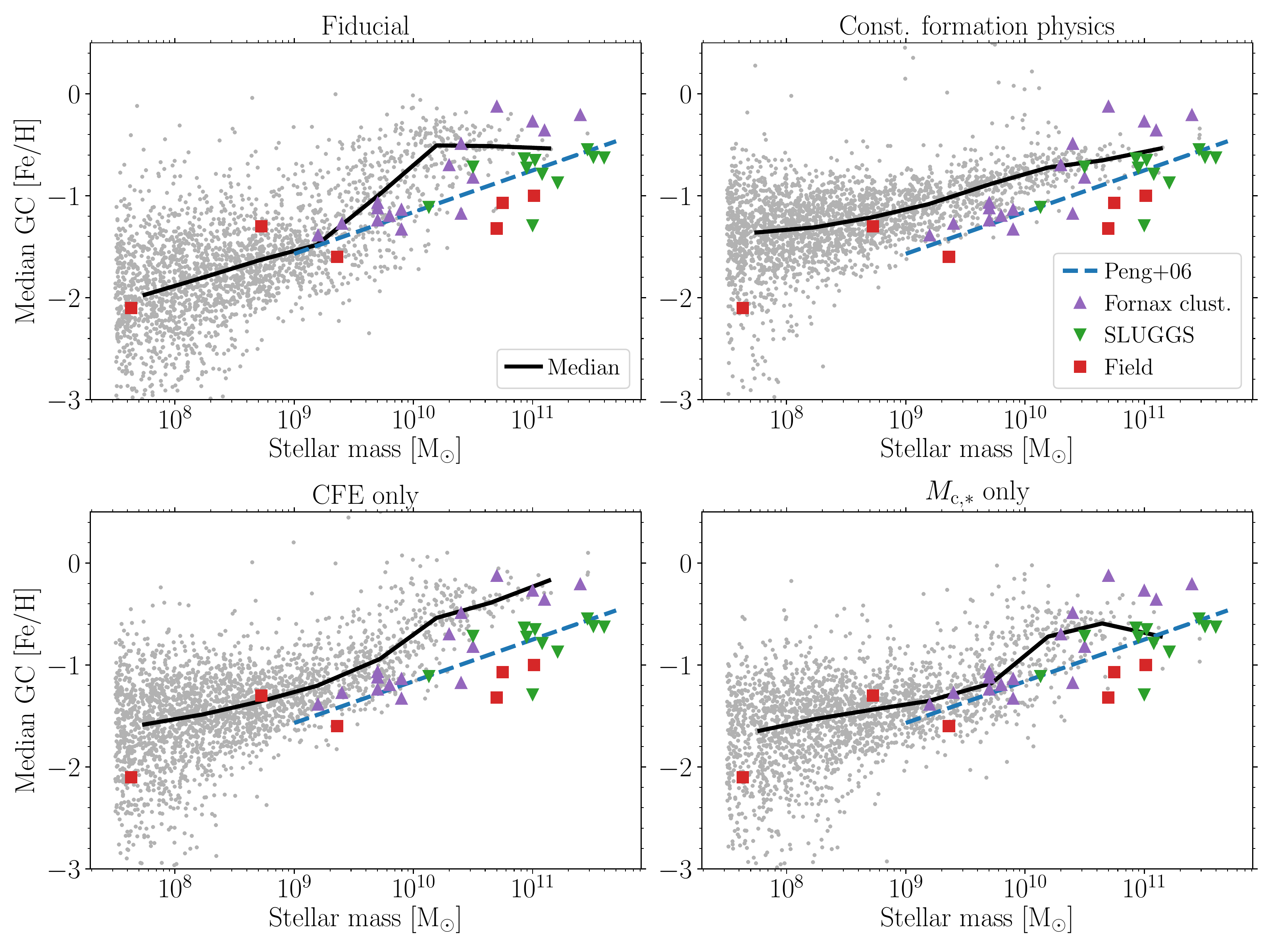}
    \caption{Median GC population metallicity as a function of galaxy mass (as in Fig.~\ref{fig:median_met}), but with a $20 \kpc$ radius limit (as in Fig.~\ref{fig:Rcut}). Symbol and line styles are identical to those in Fig.~\ref{fig:median_met}.}
    \label{fig:median_met_Rcut}
\end{figure*}

In Section~\ref{sec:alt-phys} we compare the median GC population metallicities of galaxies for the four E-MOSAICS GC formation models (Fig.~\ref{fig:median_met}).
In Section~\ref{sec:met_dists} we also show that observational aperture limits for Fornax cluster galaxies have a significant effect on the predicted GC metallicity distributions for massive galaxies (Figures~\ref{fig:met_dists} and \ref{fig:Rcut}).
Therefore, in Fig.~\ref{fig:median_met_Rcut} we repeat the comparison of median GC metallicities with a $20 \kpc$ radius limit (as in Fig.~\ref{fig:Rcut}) to demonstrate the effect of the observational aperture on the medial metallicity.
In this case, the predicted median GC metallicities for massive galaxies ($M_\ast \gtrsim 5 \times 10^{10} \Msun$) are in better agreement with Fornax/Virgo clusters and SLUGGS survey galaxies.


\bsp	
\label{lastpage}
\end{document}